\documentclass[12pt]{article}
\pdfoutput=1
\usepackage{putex}
\usepackage{comment}
\usepackage{graphicx}
\usepackage{caption}
\usepackage{amsmath}
\usepackage[usenames,dvipsnames,table]{xcolor}
\usepackage{array}
\usepackage{subcaption}
\usepackage{epstopdf}
\usepackage{enumerate}
\usepackage{cite}
\usepackage{tensor}
\usepackage{slashed}
\usepackage[export]{adjustbox}
\usepackage[aligntableaux=center]{ytableau}
\usepackage{dsfont}
\usepackage[utf8]{inputenc}
\usepackage[
      colorlinks=true,
      linkcolor=blue,
      urlcolor=blue,
      filecolor=black,
      citecolor=red,
      ]{hyperref}
\newcommand{\RNum}[1]{\uppercase\expandafter{\romannumeral #1\relax}}

\usepackage{tikz}

\newcommand{\abs}[1]{\left\lvert #1 \right\rvert}

\newcommand {\be} {\begin {equation}}
\newcommand {\ee} {\end {equation}}

\newcommand {\bes} {\begin {equation*}}
\newcommand {\ees} {\end {equation*}}

\newcommand{\R}{\mathbb{R}}

\newcommand{\cD}{{\mathcal D}}

\newcommand{\cF}{{\mathcal F}}

\newcommand{\cN}{{\mathcal N}}

\newcommand{\beq}{\begin{equation}}
\newcommand{\eeq}{\end{equation}}

\newcommand\sig{\sigma}

\def\ie{\begin{equation}\begin{aligned}}
\def\fe{\end{aligned}\end{equation}}

\numberwithin{equation}{section}

\def\<{\langle}
\def\>{\rangle}
\def \eps {\epsilon}

\begin{document}

\preprint{PUPT-2643}


\institution{PU}{Department of Physics, Princeton University, Princeton, NJ 08544, USA }

\title{\Huge Notes on a Surface Defect in the $O(N)$ Model} 

\authors{Simone Giombi, Bowei Liu}

\abstract{We study a surface defect in the free and critical $O(N)$ vector models, defined by adding a quadratic perturbation localized on a two-dimensional subspace of the $d$-dimensional CFT. We compute the beta function for the corresponding defect renormalization group (RG) flow, and provide evidence that at long distances the system flows to a nontrivial defect conformal field theory (DCFT). We use epsilon and large $N$ expansions to compute several physical quantities in the DCFT, finding agreement across different expansion methods. We also compute the defect free energy, and check consistency with the so-called $b$-theorem for RG flows on surface defects.}

\date{}
\maketitle

\tableofcontents

\section{Introduction and Summary}
\label{Sec:Introduction}

The study of boundaries and defects in quantum field theory is interesting for a
variety of reasons. From a theoretical point of view, extended objects offer a 
way to probe QFT dynamics which is complementary to that provided by local operators. From a more practical perspective, defects can be used to describe physical impurities or extended structures in quantum systems that can be realized in experiments. Recently, there has been growing interest in studying the physics of boundary and defects in conformal field theory (CFT). In this context, of particular relevance are defects that preserve a subgroup of the conformal symmetry of the ``bulk" theory. A $p$-dimensional conformal defect preserves a $SO(p + 1, 1) \times SO(d - p)$ subgroup of the conformal group $SO(d+1,1)$, and the bulk CFT coupled to the defect is referred to as a defect conformal field theory (DCFT), whose correlation functions have a rich structure constrained by the residual $SO(p + 1, 1) \times SO(d - p)$ symmetry.  In the context of the conformal bootstrap program, for instance, this setup provides new bootstrap equations that are complementary to the usual ones without the defect (see e.g. \cite{Liendo:2012hy, Billo:2016cpy}). Defects in QFT also lead to new constraints on renormalization group flow, see for instance \cite{Affleck:1991tk, Friedan:2003yc, Gaiotto:2014gha, Jensen:2015swa, Beccaria:2017rbe, Cuomo:2021rkm}.

In this paper, we study a simple example of surface defect in the $O(N)$ vector model in $d$ dimensions. The defect is defined by adding a mass-like $O(N)$ invariant term on a 2d subspace, which we take to be the $x_1, x_2$ plane (with other coordinates set to zero). The total action is  (below $\phi_a, a=1,\ldots,N$ is a $N$-component real scalar field)
\begin{equation}
S=\int d^{d} x\left[\frac{1}{2}\left(\partial \phi_{a}\right)^{2}+\frac{\lambda}{4 !}\left(\phi_{a}^{2}\right)^{2}\right]
+ h \int d x_1 d x_2\ \phi_{a}^2(\vec{x},0)\,.
\label{phi4-defect}
\end{equation}
We tune the bulk coupling constant $\lambda$ to criticality, and study the defect RG flow triggered by the defect coupling $h$. It is well-known that at the critical point of the $O(N)$ model, the operator $\phi_a^2$ has $\Delta <2$, and hence $h$ is a relevant coupling. Its beta function can be computed perturbatively in $d=4-\epsilon$ using the standard epsilon expansion techniques. Below we will perform this calculation to the leading non-trivial order in $\epsilon$, finding as expected a non-trivial $O(N)$ invariant IR fixed point of the defect RG flow. Hence, the surface defect is conformal at long distances, and the model (\ref{phi4-defect}) flows to a non-trivial DCFT with $O(N)$ symmetry. We determine the simplest scaling dimensions of local operators inserted on the defect, namely those of $\hat\phi_a$ and $\hat\phi^2_a$ operators.\footnote{In this paper, we will follow the common convention of denoting local operators on the defect with a hat, in order to distinguish them from the bulk operators, which have different scaling dimensions.} We also study the same system using the large $N$ expansion in general $d$, as well as the epsilon expansion in the cubic theory \cite{Fei:2014yja} near $d=6$, finding agreement between these different descriptions in their overlapping regime of validity. In $d=4-\epsilon$, we also compute perturbatively the free energy of the model for the case where the defect is supported on a two-sphere, and find a result consistent with the so-called $b$-theorem \cite{Jensen:2015swa} for 2d defect RG flow.

The planar defect (\ref{phi4-defect}) was previously mentioned in \cite{ParisenToldin:2016szc} in the context of the critical behavior of spin systems, but it was not studied in detail there. More recently, it was studied in \cite{Krishnan:2023cff} for $d=3$. In this case, the defect becomes an interface and the problem is closely related to that of the $O(N)$ model in the presence of a boundary. This problem has recently received renewed attention, in particular in connection to the so-called ``extraordinary-log" phase proposed in \cite{Metlitski:2020cqy}, see also \cite{Padayasi:2021sik, ParisenToldin:2016szc}.  Other recent related work includes applications to quantum Hall bilayers \cite{zhang:2023xy} and weak measurement systems \cite{lee2023:quantum}. 

While the analysis of \cite{Krishnan:2023cff} focuses on $d=3$ where the defect is an interface, here we will keep the dimension of the ``bulk" CFT to be general $d$, and the defect to be always two-dimensional. Interestingly, we find that the results of the perturbative large $N$ expansion in general $d$ cannot be directly continued to $d=3$, as they naively diverge in the $d\rightarrow 3$ limit. This suggests a qualitatively different behavior of the large $N$ theory in $d=3$ and $3<d<4$, which we discuss in more detail in Section \ref{largeN} below.  

In this paper we focus mostly on the $O(N)$ invariant fixed point of (\ref{phi4-defect}), but the theory is also expected to admit a phase where the $O(N)$ symmetry is broken, 
analogously to the case of the $O(N)$ model in the presence of a boundary (see \cite{Diehl:1996kd} for a review). 
In this phase, the bulk operator $\phi_a$ acquires a non-trivial one-point function growing towards the defect. We briefly discuss such $O(N)$ breaking phase in Section \ref{sym_4} below, both using the equations of motion in flat space, and by mapping the theory to $H^3\times S^{d-3}$. Here $H^3$ is the three-dimensional hyperbolic space, with the surface defect being at its boundary. In this case, the $O(N)$ breaking phase can be understood as a critical point of the classical potential in hyperbolic space, similar to the BCFT setup discussed in \cite{Giombi:2020rmc}. 


The rest of this paper is organized as follows: as a simple example, in section \ref{sec:free} we first study the model (\ref{phi4-defect}) in the case where we tune the bulk coupling to the free UV fixed point ($\lambda=0$), so that we have a planar defect in the theory of $N$ free massless scalar fields.\footnote{This example was also recently discussed in \cite{Shachar:2022fqk}.} We show that the beta function for $h$, as well as the scaling dimensions of defect operators, can be computed exactly in this case. We also discuss the mapping to $H^3\times S^{d-3}$ and point out a close relation to the problem of double-trace deformations in AdS/CFT \cite{Klebanov:1999tb, Gubser:2002vv, Diaz:2007an}. Then in Section \ref{sec:4-eps} we use the epsilon expansion to study the planar defect in the interacting $O(N)$ model in $d=4-\epsilon$, and in Section \ref{sec:largeN} we study the system using the large $N$ expansion in generic $d$. As a further cross-check of the large $N$ results, in Section \ref{sec:6-eps} we study the planar defect in the cubic scalar theory in $6-\eps$ expansion. Finally in section \ref{sec:F} we compute the free energy for the spherical defect first in the case of the free theory (exactly) and then for the interacting theory in $d=4-\epsilon$ (perturbatively in $\epsilon$), finding results in agreement with the defect $b$-theorem \cite{Jensen:2015swa}.  

\paragraph{Note added.} While writing up this paper, we became aware of  \cite{Raviv-Moshe:2023yvq} and \cite{Trepanier:2023tvb}, which present results that overlap with parts of this work. We thank Avia Raviv-Moshe and Siwei Zhong for informing us of their work \cite{Raviv-Moshe:2023yvq} and for sharing a preliminary draft prior to submission. 

\section{Free Theory}
\label{sec:free}
Let us first discuss the surface defect in the case of the free scalar field theory. We consider $N$ free scalar fields in $d$ dimension, $\phi_a$, $a=1,2,\cdots,N$, and insert a $O(N)$ invariant surface defect onto the $x_1, x_2$ plane. The action is
\begin{equation}
S=\int d^{d} x\ \frac{1}{2}\left(\partial \phi_{a}\right)^{2}
+ h_{0} \int d x_1 d x_2\ \phi_{a}^2
\label{free_action}\end{equation}
Of course, since the theory is free, the $N$ dependence is trivial and one might as well restrict to a single scalar. However, since we later generalize to the case of the interacting $O(N)$ model, we will work with $N$ scalars in this section. Recall that the free scalar propagator in $d$ dimensions is given by
\begin{equation}
G(x-y)=\int \frac{d^d p}{(2 \pi)^d} \frac{e^{i p(x-y)}}{p^2}
= \frac{C_\phi}{|x-y|^{d-2}}\,,\qquad C_{\phi}\equiv \frac{\Gamma \left(\frac{d-2}{2}\right)}{4\pi^{\frac{d}{2}}} 
\label{prop}
\end{equation}

The operator $\phi_a^2$ inserted on the 2d defect is a relevant deformation in $d<4$, and it is expected to trigger a defect RG flow. The usual perturbative renormalization can be developed in $d=4-\epsilon$, where the beta function of the defect coupling can be obtained by familiar techniques. Explicitly, to renormalize the defect coupling $h_0$, we require the one-point function $\< \phi_a^2 (0, x)\>$ at a distance $\abs{x}$ from the defect to be finite as $\epsilon\rightarrow0$. The diagrams contributing to $\<\phi_a^2 (0, x)\>$ form an infinite series that we may depict as follows 
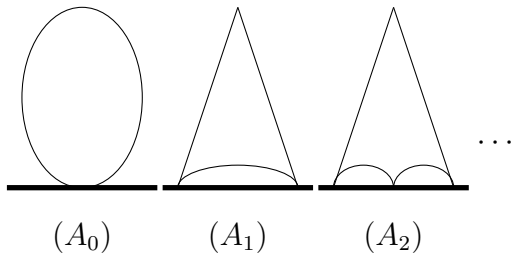
\begin{equation*}
\begin{gathered}
\begin{tikzpicture}
\coordinate[label=below:{$(A_0)$}] (A) at (0,-1.5);
\draw (0,0) ellipse (0.8 and 1.2);
\draw[line width=2] (-1, -1.2) -- (1, -1.2);
\end{tikzpicture}
\begin{tikzpicture}
\coordinate[label=below:{$(A_1)$}] (A) at (0,-1.5);
\draw (0, 1.2) -- (-0.8, -1.2);
\draw (0, 1.2) -- (0.8, -1.2);
\draw (0.8, -1.2) arc(0:180:0.8 and 0.3);
\draw[line width=2] (-1, -1.2) -- (1, -1.2);
\end{tikzpicture}
\begin{tikzpicture}
\coordinate[label=below:{$(A_2)$}] (A) at (0,-1.5);
\draw (0, 1.2) -- (-0.8, -1.2);
\draw (0, 1.2) -- (0.8, -1.2);
\draw (0, -1.2) arc(0:180:0.4 and 0.3);
\draw (0.8, -1.2) arc(0:180:0.4 and 0.3);
\draw[line width=2] (-1, -1.2) -- (1, -1.2);
\end{tikzpicture}
\end{gathered}
\cdots
\end{equation*}
where all the defect vertices form an infinite chain.

It will be more convenient to implement the renormalization in the momentum space representation. We may write 
\be
\< \phi_a^2 (0, x)\>
= \int \frac{d^{d-2} m} {(2\pi)^{d-2}} \int \frac{d^{d-2} n} {(2\pi)^{d-2}} \int \frac{d^2 k}{(2\pi)^2}
e^{-i(m+n)x} \langle \phi_a (k, m) \phi_a (-k, n) \rangle\,.
\ee
Here we denote the fields in momentum space by $\phi_a(k,m)$, where $k$ is a two-dimensional momentum along the defect (which is conserved), and $m$ is $(d-2)$-dimensional momentum. To renormalize the bulk one-point function, we may then equivalently require that $\langle \phi_a (k, m) \phi_a (-k, n) \rangle$ is finite as $\epsilon\rightarrow0$.
The leading contribution to $\<\phi_a (k, m) \phi_a (-k, n)\>$ is given by
\begin{equation}
{A_0} = - 2 h_0 \frac{N}{(k^2+m^2) (k^2+n^2)}
\end{equation}
Since the 2-dimensional defect momentum $k$ is conserved, insertion of a new defect vertex gives a factor of
\begin{equation}
t(k)= -2h_0 \int \frac{d^{d-2} p}{(2\pi)^{d-2}} \frac{1}{k^2+p^2}
= -h_0 2^{3-d} \pi ^{1-\frac{d}{2}} k^{d-4} \Gamma \left(2-\frac{d}{2}\right)
\end{equation}
Insertion of more vertices gives a power of $t(k)$. In total, 
\begin{equation}
A_n = t^n (k) A_0
\end{equation}
\begin{equation}
\<\phi_a (k, m) \phi_a (-k, n)\> = \sum_{n=0}^\infty A_n = \frac{A_0}{1-t(k)} 
= \frac{- 2 h_0 \frac{N}{(k^2+m^2) (k^2+n^2)}}{1+h_0 2^{3-d} \pi ^{1-\frac{d}{2}} k^{d-4} \Gamma \left(2-\frac{d}{2}\right)}
\end{equation}
At $d=4-\eps$, $t(k) = - \frac{h_0}{\pi \eps} +O\left(\epsilon ^0\right)$. So if we define
\begin{equation}\label{h}
\frac{M^{\epsilon}}{h_0} = \frac{1}{h} -\frac{1}{\pi \epsilon}, \quad h_0 = M^{\epsilon} \frac{h}{1-h/(\pi \epsilon)} 
\end{equation}
where $M$ is a renormalization scale, then
\begin{equation}
\begin{aligned}
\<\phi_a (k, m) \phi_a (-k, n)\> 
=& \frac{-2 h M^\eps}{(k^2+m^2) (k^2+n^2)} \frac{N}{1 - \frac{h}{\pi\eps} + h {M^\eps} 2^{3-d} \pi ^{1-\frac{d}{2}} k^{d-4} \Gamma \left(2-\frac{d}{2}\right)} \\
=& \frac{-2 h M^\eps}{(k^2+m^2) (k^2+n^2)} \frac{N}{1+\frac{h}{2 \pi} \left(\log \left(\frac{4 \pi  M^2}{k^2}\right)-\gamma +O\left(\epsilon ^1\right) \right)}
\end{aligned}
\end{equation}
is finite. 

Taking the derivative of eq.~(\ref{h}) w.r.t.~$M$ and requiring the derivative of the bare coupling $h_
0$ vanishes $\left({\partial h_0}/{\partial M} = 0\right)$, we arrive at the beta function 
\begin{equation}
\beta_h = - \epsilon h + \frac{h^2}{\pi}
\label{free_beta}\end{equation}
Note that this beta function is {\it exact}, since the definition of renormalized coupling obtained above cancel divergences to all orders in $h$. 
Then we find an exact IR fixed point at
\begin{equation}\label{free_fp}
h_* = \pi \epsilon = \pi (4-d)
\end{equation}

\paragraph{Conformal perturbation theory approach.}\label{sec:cp}
The beta function obtained above by renormalizing the bulk one-point function $\langle \phi^2_a\rangle$ can also be recovered by using the standard conformal perturbation theory approach, adapted to the 2-dimensional defect. In general, if we consider a CFT perturbed by 
a weakly relevant operator $\hat{O}$ on a 2d defect, with scaling dimension ${\Delta}=2-\varepsilon$, 
\begin{equation}
S = S_{\mathrm{CFT}}+{h}_0 \int {d}^2 {x} \sqrt{{g}} \hat{O}(x)
\end{equation}
then the beta function for the dimensionless renormalized coupling $h$ is~\cite{Kobayashi:2018lil} (this is a straightforward adaptation of the 
general conformal perturbation theory result in a CFT perturbed by a weakly relevant operator, see e.g. \cite{Cardy:1988cwa, Klebanov:2011gs, Fei:2015oha})
\begin{equation}
\beta_{h}=-\varepsilon {h}+ \pi \frac{C_3}{C_2} {h}^2+O\left({h}^3\right)
\end{equation}
where $C_3 = \langle \hat{O} \hat{O} \hat{O}\rangle_0$ and $C_2 =\langle \hat{O} \hat{O}\rangle_0$ are the 3-point and 2-point function normalizations (omitting the standard position dependent factors) in the unperturbed CFT.  
Applied to our model in eq.~(\ref{free_action}), $\hat{O} = \hat{\phi}_a^2$ and $\varepsilon = \epsilon$, so we have
\begin{equation}
\frac{C_3}{C_2} = \frac{8 N C_\phi^3}{2N C_{\phi}^2} = 4 C_{\phi}
\label{C-coeffs}
\end{equation}
Therefore
\begin{equation}
\beta_{h} = - \eps h + 4 \pi C_\phi h^2 + O(h^3) 
\end{equation}
Note that in $d=4-\eps$ to leading order in $\epsilon$, we have $4 \pi C_\phi=1/\pi$. So this beta function is consistent as expected with eq.~(\ref{free_beta}) computed in the minimal subtraction scheme.\footnote{In general, the beta function in minimal subtraction and that obtained in conformal perturbation theory can be thought of as corresponding to two different schemes. Observables at the fixed point, such as scaling dimensions, are of course independent of the renormalization scheme.} However, from the conformal perturbation theory approach it is not obvious that the beta function is in fact exact, as we have shown explicitly above by renormalizing the bulk one-point function and summing up all diagrams.  

\subsection{Scaling dimensions on the defect}
We now calculate the scaling dimensions of the operators $\hat{\phi}_a$ and $\hat{\phi}^2$ inserted on the defect. 

For the operator $\hat{\phi}_i$, let us consider the 2-point function $\langle \hat{\phi}_1(y_1) \hat{\phi}_1(y_2)\rangle_d$ where $y_1, y_2$ are coordinates on the two-dimensional defect plane. If we do a two-dimensional Fourier transform along the defect plane (here $p_1, p_2$ are two-dimensional defect momentum), we have
\begin{equation}\label{defect_pro}
\begin{aligned}
&\langle\hat{\phi}_i(p_1) \hat{\phi}_j(p_2)\rangle
= \int d^2 y_1 \int d^2 y_2 \ e^{i(p_1 y_1 + p_2 y_2)} 
\frac{C_\phi}{|y_1 - y_2|^{d-2}} \delta_{ij} \\
=& \int d^2 y_1 \int d^2 \Delta y \ e^{i(p_1 y_1 + p_2 (\Delta y + y_1))} 
\frac{C_\phi}{|\Delta y|^{d-2}} \delta_{ij} \\
=& (2\pi)^2 \delta^{(2)}(p_1 + p_2)
\frac{2^{2-d} \pi ^{1-\frac{d}{2}} \Gamma \left(2-\frac{d}{2}\right)}{|p_2|^{4-d}} \delta_{ij} \\
=& (2\pi)^2 \delta^{(2)}(p_1 + p_2)
\frac{t(p_2)}{-2h_0} \delta_{ij} \,.
\end{aligned}
\end{equation}
The defect action in momentum space is simply
\begin{equation}
S = h_0 \int \frac{d^2 k}{(2\pi)^2}\,.
\hat{\phi}_i (k) \hat{\phi}_i (-k) 
\end{equation}
Because the defect action is quadratic, every diagram in $\langle\hat{\phi}_i(p_1) \hat{\phi}_j(p_2)\rangle_D$ is a chain on the defect:
\begin{equation}
\begin{gathered}
\begin{tikzpicture}
\draw (0.8, -1.2) arc(0:180:0.8 and 0.3);
\draw[line width=2] (-1, -1.2) -- (1, -1.2);
\end{tikzpicture}
\begin{tikzpicture}
\draw (0, -1.2) arc(0:180:0.4 and 0.3);
\draw (0.8, -1.2) arc(0:180:0.4 and 0.3);
\draw[line width=2] (-1, -1.2) -- (1, -1.2);
\end{tikzpicture}
\end{gathered}
\cdots
\label{chain}
\end{equation}
Momentum is conserved and the diagrams form a power series:
\begin{equation}
\langle\hat{\phi}_i(p_1) \hat{\phi}_j(p_2)\rangle_{D} 
= \sum_{n=0}^\infty (2\pi)^2 \delta^{(2)}(p_1 + p_2) \delta_{ij} 
\frac{t(p_2)}{-2h_0} t^n (p_2) 
\end{equation}
Because $t(p) = -h_0 2^{3-d} \pi ^{1-\frac{d}{2}} p^{d-4} \Gamma \left(2-\frac{d}{2}\right)$ is divergent as $\eps\rightarrow0$, the leading order diagram is divergent. To avoid this issue, it is more convenient to perform the renormalization in terms of the coordinate space representation.
We can write 
\begin{equation}
\langle \hat{\phi}_1 (y_1) \hat{\phi}_1 (y_2)\rangle_D = \sum_{n=0}^\infty C_n
\end{equation}
where, computing the Fourier transform of $\frac{t^{n+1}(p_2)}{-2h_0}$ by using eq.~(\ref{fourier}), we get
\begin{equation}
C_n
= -\frac{2^{(d-4) n+d-5} |y_1-y_2|^{-d (n+1)+4 n+2} \Gamma \left(\frac{1}{2} (d+(d-4) n-2)\right) \left(-2^{3-d} \pi ^{1-\frac{d}{2}} h_0 \Gamma
   \left(2-\frac{d}{2}\right)\right)^{n+1}}{\pi  h_0 \Gamma \left(-\frac{1}{2} (d-4) (n+1)\right)}
\end{equation}
We can define the renormalized operator as usual by
\begin{equation}
\hat{\phi}_i (y)=Z_{\hat{\phi}}\left[\hat{\phi}_i \right](y)
\end{equation}
where $[\hat{\phi}_i]$ denotes the renormalized operator on the defect. Then we substitute this and eq.~(\ref{h}) into the expression for $C_n$, and require that
it is finite. This gives 
\begin{equation}
Z_{\hat{\phi}} = 1 - \frac{h}{\pi\epsilon}
\end{equation}
The anomalous dimension of $\hat{\phi}_i$ is then
\begin{equation}
\gamma_{\hat{\phi}} = \frac{\partial \log Z_{\hat{\phi}}}{\partial \log M} = \beta_h \frac{\partial \log Z_{\hat{\phi}}}{\partial h} 
= \frac{-1}{\pi\epsilon Z_{\hat{\phi}}} ( - \epsilon h + \frac{h^2}{\pi}) 
= \frac{h}{\pi}
\end{equation}
At the fixed point $h_* = \pi\epsilon$, this gives
\begin{equation}
\Delta_{\hat{\phi}} = \frac{d-2}{2} +\left.\gamma_{\hat{\phi}}\right\vert_{h=h_*} = 1 + \frac{\epsilon}{2} = 3-\frac{d}{2}
\label{free-delphi}
\end{equation}
where in the last step we expressed the scaling dimension in terms of $d$, because the result is exact in $\epsilon$. 

As the defect action is $h_{0} \int d x_1 d x_2\, \phi_{a}^2$, the scaling dimension of the operator $\hat{\phi}^2_a$ inserted on the defect can be calculated directly from the derivative of the beta function at the fixed point.  
Using the beta function and fixed point in eq.~(\ref{free_beta}) and~(\ref{free_fp}), we find
\begin{equation}\label{delta_phi2_f}
\Delta_{\hat{\phi}^2} = 2 + \left.\frac{\partial \beta_h}{\partial h}\right\vert_{h=\pi\eps} 
= 2 + \eps = 6-d\,.
\end{equation}
Note that this satisfies $\Delta_{\hat{\phi}^2} = 2\Delta_{\hat{\phi}}$. This is expected since, even though we have a non-trivial defect fixed point, the theory is Gaussian. 

Note that in the special case $d=3$, we have $\Delta_{\hat{\phi}} = \frac{3}{2}$. This is the dimension of the boundary scalar operator in the 3d free scalar theory with Dirichlet boundary conditions on a 2d boundary. This is as expected, since for $d=3$ our defect becomes an interface, and the IR fixed point should correspond to two copies of the Dirichlet free scalar BCFT. This can also be confirmed by computing the defect free energy (and related defect anomaly coefficient), as we discuss in Section \ref{sec:F} below. 

In general $d$, a simple way to understand the result (\ref{free-delphi}) is by performing a Weyl transformation to the space $H^3\times S^{d-3}$, where $H^3$ is the three-dimensional hyperbolic space. The flat space metric can be written as
\begin{equation}
\begin{aligned}
& ds^2 = dx_1^2+dx_2^2+dr^2 +r^2 d\Omega_{d-3}^2 \\
=& r^2 [ \frac{dx_1^2+dx_2^2+dr^2}{r^2} + d\Omega_{d-3}^2]
=r^2 ds^2_{H^3 \times S^{d-3}}
\label{metric-weyl}
\end{aligned}
\end{equation}
By making a Weyl rescaling to get rid of the overall conformal factor, we can then map the DCFT on flat space to $H^3 \times S^{d-3}$, where the 2d defect now sits at the boundary of $H^3$. Including the conformal coupling term, the bulk action becomes (see for instance \cite{Giombi:2020rmc})
\begin{equation}
S=\int d^{d} x \sqrt{g} \left[\frac{1}{2}\left(\partial \phi_{a}\right)^{2}
+\frac{(d-2)(d-6)}{8} \phi_a^2\right]
\end{equation}
If we perform a Kaluza-Klein reduction on $S^{d-3}$, we obtain a tower of states with masses $m_{\ell}^2 =\frac{1}{4}(d-2)(d-6) + \ell(\ell+d-4)$, where $\ell=0,1,2,\ldots$ and we used the standard result for the eigenvalues of the Laplacian on the sphere $S^{d-3}$. For the lowest lying mode, we have $m^2_0 = \frac{1}{4}(d-2)(d-6)$, which according to the mass/dimension relation on $H^3$ given by $\Delta(\Delta-2) = m^2_0$, gives the two possible operator dimensions at the boundary of $H^3$:
\begin{equation}
\Delta_-  =\frac{d}{2}-1 \,\qquad \Delta_+ =3-\frac{d}{2} 
\end{equation}
We see that these are indeed respectively the dimensions of the defect operator $\hat{\phi}_a$ at the trivial fixed point ($h=0$) and at the IR fixed point ($h=h_*$). From this point of view, we can think of the surface defect flow in the theory (\ref{free_action}) as an analog of the familiar double-trace flow in the AdS/CFT context \cite{Klebanov:1999tb, Gubser:2002vv, Diaz:2007an}. Here, the analog of the double-trace deformation is the operator $\hat{\phi}_a^2$ inserted at the boundary of $H^3$.

\section{Interacting $O(N)$ model in $d=4-\epsilon$}
\label{sec:4-eps}
We now consider the massless $O(N)$ symmetric Wilson-Fisher model in $d=4-\eps$ dimension with the same surface defect inserted on the $x_1, x_2$ plane:
\begin{equation}
S=\int d^{d} x\left[\frac{1}{2}\left(\partial \phi_{a}\right)^{2}+\frac{\lambda_{0}}{4 !}\left(\phi_{a}^{2}\right)^{2}\right]
+ h_{0} \int d x_1 d x_2\ \phi_{a}^2
\label{phi4}\end{equation}

Since in the free theory the fixed point of the defect coupling is at $h_* = \pi \epsilon$, it is natural to expect that at the interacting bulk fixed point, $h_* = O(\eps)$ as well. Then both $\lambda_*$ and $h_*$ are of order $O(\eps)$, and there are three diagrams to two-loop order of $h_0$ and $\lambda_0$:
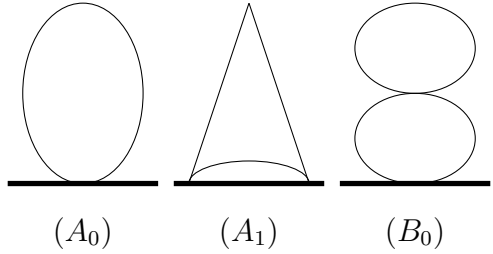
\begin{center}
\begin{tikzpicture}
\coordinate[label=below:{$(A_0)$}] (A) at (0,-1.5);
\draw (0,0) ellipse (0.8 and 1.2);
\draw[line width=2] (-1, -1.2) -- (1, -1.2);
\end{tikzpicture}
\begin{tikzpicture}
\coordinate[label=below:{$(A_1)$}] (A) at (0,-1.5);
\draw (0, 1.2) -- (-0.8, -1.2);
\draw (0, 1.2) -- (0.8, -1.2);
\draw (0.8, -1.2) arc(0:180:0.8 and 0.3);
\draw[line width=2] (-1, -1.2) -- (1, -1.2);
\end{tikzpicture}
\begin{tikzpicture}
\coordinate[label=below:{$(B_0)$}] (A) at (0,-1.5);
\draw (0,0.6) ellipse (0.8 and 0.6);
\draw (0,-0.6) ellipse (0.8 and 0.6);
\draw[line width=2] (-1, -1.2) -- (1, -1.2);
\end{tikzpicture}
\end{center}

In position space, these diagrams can be computed as
\begin{equation}\label{A0}
A_0 = - 2 h_0 N \int d^2 y
\frac{C_\phi^2}{(x^{2}+y^{2})^{(d-2)}}
= \frac{\pi ^{1-d} h_0 N x^{6-2 d} \Gamma \left(\frac{d}{2}-1\right)^2}{24-8 d}
\quad\text{eq.~\ref{I3}}
\end{equation}
\begin{equation}
\begin{aligned}
{A_1} =&
4 h_0^2 N \int d^2 y \int d^2 z
\frac{C_\phi^3}{(x^{2}+y^{2})^{(d-2)/2} (x^{2}+z^{2})^{(d-2)/2} |y-z|^{(d-2)}} \\
=& -\frac{\pi ^{3-\frac{3 d}{2}} h_0^2 N x^{10-3 d} \csc \left(\frac{\pi  d}{2}\right) \Gamma (d-3)^2 \Gamma \left(\frac{3 d}{2}-5\right)}{16 \Gamma (2 d-6)}
\quad\text{eq.~\ref{I38}}
\end{aligned}
\label{A1}
\end{equation}
\begin{equation}
\begin{aligned}
B_0 =& \frac{(N+2) \lambda_0 h_0 N}{3} \int d^2 y \int d^d z
\frac{C_\phi^4}{|z-y|^{2(d-2)} |z-x|^{2(d-2)}} \\
=& \frac{\pi ^{3-\frac{3 d}{2}} h_0 \lambda_0 N (N+2) x^{10-3 d} \csc ^2\left(\frac{\pi 
   d}{2}\right) \Gamma \left(\frac{d}{2}-1\right)^2 \Gamma \left(\frac{3
   d}{2}-5\right)}{768 \Gamma (4-d) \Gamma (d-2)^2}
\end{aligned}\label{B0}
\end{equation}

The wavefunction renormalization factor is the same as the one in the bulk theory without defect (see for instance \cite{kleinert2001critical} for a collection of CFT data 
of the $O(N)$ model in $d=4-\epsilon$):
\begin{equation}
\phi_{a}^{2}(x)=Z_{\phi^{2}}\left[\phi_{a}^{2}\right](x),
\quad
Z_{\phi^{2}}=1-\frac{\lambda(N+2)}{48 \pi^{2} \epsilon}+\mathcal{O}\left(\frac{\lambda^{2}}{(4 \pi)^{4}}\right)
\label{Z_phi^2b}\end{equation}
The renormalized one-point function $\langle [\phi_a^2] (x)\rangle$ is
\begin{equation}
\begin{aligned}
&\langle [\phi_a^2] (x)\rangle
= \frac{1}{Z_{\phi^2}} (A_0 + A_1 + B_0)\,.
\end{aligned}
\end{equation}
If we now substitute in the renormalized couplings
\begin{equation}\label{lambda0}
\lambda_{0}=M^{\epsilon}\left(\lambda+O(\lambda^2)\right)
\end{equation}
\begin{equation}
h_0 = M^{\epsilon} \left(\frac{h}{1-h/(\pi \epsilon)}
+ \lambda \frac{a_{11}}{\epsilon} h \right)
\end{equation}
and require the renormalized one-point function $\langle [\phi_a^2] (x)\rangle$ to be finite, we find
\begin{equation}
h_0 = M^{\epsilon} \left(\frac{h}{1-h/(\pi \epsilon)}
+ \lambda \frac{N+2}{48\pi^2 \epsilon} h\right)\,.
\label{h0}\end{equation}
Imposing $\frac{\partial h_0}{\partial \log(M)} = 0$, we arrive at the beta function
\begin{equation}\label{betah:lambdah}
\beta_h = -h \epsilon + \frac{h^2}{\pi }+\frac{\lambda h (N+2)}{48 \pi ^2}
\end{equation}
This corresponds to a fixed point at
\begin{equation}
h_* = \pi\epsilon -\frac{(N+2)\lambda}{48 \pi }
\end{equation}
Plugging in the explicit value of the bulk fixed point coupling
\begin{equation}\label{lambda*}
\frac{\lambda_{*}}{(4 \pi)^{2}}=\frac{3 \epsilon}{N+8}+\frac{9(3 N+14) \epsilon^{2}}{(N+8)^{3}}+\mathcal{O}\left(\epsilon^{3}\right)\,,
\end{equation}
the defect coupling fixed point is
\begin{equation}\label{h*:lambdah}
h_* = \frac{6 \pi}{N+8} \epsilon + \mathcal{O}\left(\epsilon^2\right)
\end{equation}
This confirms our assumption that $h_* = O(\eps)$ and shows that our calculation is self-consistent.

At this fixed point, the one-point function of the bulk quadratic operator is
\begin{equation}
\langle [\phi_a^2] (x)\rangle 
= -\frac{3 \epsilon N}{4 \pi ^2 (N+8) x^2}+O\left(\epsilon^2\right)
\label{one-point-phisq}
\end{equation}
To compare with the results of large $N$ expansion in Section~\ref{largeN}, it will be useful to normalize the one-point function by the square root of the two-point function coefficient. Since
\begin{equation}
\left\langle\left[\phi_{a}^{2}\right](x)\left[\phi_{b}^{2}\right](0)\right\rangle=\frac{\mathcal{N}_{\phi^{2}}^{2}}{|x|^{2\left(d-2+\gamma_{\phi^{2}}\right)}}
\end{equation}
where (see e.g. \cite{Cuomo:2021kfm})
\begin{equation}
\mathcal{N}_{\phi^{2}}^{2}=\frac{2 N}{(d-2)^{2} \Omega_{d-1}^{2}}\left[1-\epsilon \frac{(N+2)\left(\gamma_{E}+1+\log \pi\right)}{N+8}+\mathcal{O}\left(\epsilon^{2}\right)\right]\,,
\end{equation}
the normalized one-point function is
\begin{equation}
\frac{\langle[\phi_a^2](x)\rangle_D} {\cN_{\phi^2}}
= -\frac{3 \epsilon \sqrt{N} }{\sqrt{2} (N+8) x^2}+O\left(\epsilon ^2\right)
\label{phi2_4n}\end{equation}

\paragraph{Conformal perturbation theory approach.} Now we use conformal perturbation theory as in Section~\ref{sec:cp} as an alternative way to obtain the beta function.
The interacting CFT in the bulk has the renormalized coupling $\lambda_*$ given by eq.~(\ref{lambda*}). We perturb it by the defect term $h_{0} \int d x_1 d x_2 \phi_{a}^2$.
The three-point and two-point function coefficients $C_3$ and $C_2$ in the unperturbed interacting CFT are the same as in free theory (eq.~\ref{C-coeffs}) to leading order. But now the scaling dimension of $\phi_a^2$ in the CFT at $h=0$ is given by the well-known result \cite{kleinert2001critical}
\begin{equation}
\Delta_{\phi^2}
= 2 - \eps + \gamma_{\phi^2}
= 2 - \frac{6 \epsilon}{N+8}+O(\epsilon^2)\,.
\end{equation}
Therefore, the dimension of the perturbing defect operator is $2-\varepsilon$ with $\varepsilon=2-\Delta_{\phi^2} = \frac{6\eps}{N+8}$, and the beta function is given by
\begin{align}
\beta_{h}
=& - \frac{6\eps}{N+8} h + 4 \pi C_\phi h^2 + O(h^3) \\
=& - \frac{6\eps}{N+8} h + \frac{h^2}{\pi} + O(h^3)
\end{align}
where in the last line we have replaced $4\pi C_{\phi}$ by its leading order value near $d=4$. This indeed agrees with eq.~(\ref{betah:lambdah}) when $\lambda=\lambda_*$ is at the critical point.

\subsection{Scaling dimensions on the defect}
Similar to the calculation for free scalar fields, we consider $\langle \hat{\phi}_1(y_1) \hat{\phi}_1(y_2)\rangle_D$ where $y_1, y_2$ are coordinates on the two-dimensional defect plane. To the linear order in $\epsilon$, the relevant diagrams are the same as in the free theory, namely we just need the diagram linear in $h$ in the ``chain" eq.~\ref{chain}. There are no diagrams involving the bulk coupling to this order, since these first appear at order $\lambda^2$ and $h\lambda$.  
Hence, to the leading non-trivial order, the $Z$ factor and anomalous dimension as functions of the renormalized coupling $h$ are the same as in the free theory
\begin{equation}
\begin{aligned}
&Z_{\hat{\phi}} = 1 - \frac{h}{\pi  \epsilon } + \frac{h \lambda  (N+2)}{96 \pi ^3} \left(-\frac{1}{\epsilon^2} + \frac{1}{\epsilon}\right)\,,\\
&\gamma_{\hat{\phi}} = \frac{h}{\pi}+O(h\lambda, \lambda^2)
\end{aligned}
\end{equation}

At the fixed point $h_* = \frac{6 \pi}{N+8} \epsilon$, this gives the scaling dimension
\begin{equation}
\Delta_{\hat{\phi}} = \frac{d-2}{2} +\left.\gamma_{\hat{\phi}}\right\vert_{h=h_*, \lambda=\lambda_*} = 1 - \frac{\epsilon}{2} + \frac{6 \epsilon }{N+8} + O(\epsilon^2)\,.
\label{Deltaphi}
\end{equation}

For the dimension of the $\hat{\phi}^2$ defect operator, we can again simply use the general formula relating it to the derivative of the beta function at the fixed point. This gives 
\begin{equation}\label{delta_phi2_4}
\Delta_{\hat{\phi}^2} = 2 + \left.\frac{\partial \beta_h}{\partial h}\right\vert_{h=h_*, \lambda=\lambda_*} 
= 2 + \frac{6 \epsilon}{N+8}\,.
\end{equation}
We will see below that these results are consistent with the predictions of the large $N$ expansion. Note that in $d<4$ the operator $\hat\phi^2$ is irrelevant at the fixed point, as expected for a IR stable fixed point. 

\subsection{$O(N)$ symmetry breaking phase}
\label{sym_4}
In the above calculation of the beta function, we found a $O(N)$ invariant DCFT where the fixed point coupling $h_*$ is positive and the flow is perturbative for small $\epsilon$. 
By analogy with the similar problem of the ``extraordinary" and ``normal" transitions in the $O(N)$ model in the presence of a boundary (see for instance \cite{Liendo:2012hy,Giombi:2020rmc}, and  \cite{Diehl:1996kd} for a review), it is natural to also look for a phase of the surface defect theory where the $O(N)$ symmetry is broken to $O(N-1)$.\footnote{More general pattern of symmetry breaking should also be possible, but here we focus on the simplest breaking to $O(N-1)$.} As in the boundary case, one can describe this phase by finding a classical solution to the equations of motion where one of the scalars, say $\phi_N$, has a non-zero classical profile, growing towards the location of the defect. One way to reach this phase is by adding an explicit $O(N)$ breaking relevant perturbation on the defect, proportional to $\phi_N$. This is the analog of the so-called ``normal" transition in the $O(N)$ model with a boundary. It is natural to expect that an $O(N)$ breaking phase can also be reached by starting with the defect action (\ref{phi4}) and taking the coupling $h$ to be negative (this is the analog of the ``extraordinary" transition in the boundary problem, corresponding to spontaneous breaking). Here we will not discuss in detail the difference between these two setups (see \cite{Metlitski:2020cqy,Toldin:2021kun, Krishnan:2023cff}), and for simplicity in the discussion below we will have in mind the analog of the ``normal" transition.

Let us place the defect on the $x_1, x_2$ plane as before, and use the coordinates
\begin{equation}
ds^2 = dx_1^2+dx_2^2+dy_1^2+ \ldots +dy_{d-2}^2\,.
\end{equation}
Then the one point function of a bulk scalar operator in the presence of the surface defect must take the form
\begin{equation}
\<O\> = \frac{c}{r^{\Delta}}
\label{one-point}
\end{equation}
where $r = \sqrt{y_1^2+y_2^2+...+y_{d-2}^2}$ is the transverse distance from the defect and $\Delta$ is the dimension of the bulk operator. For $d=4-\eps$, we can solve for the equation of motion (as in the boundary case \cite{Liendo:2012hy})
\begin{equation}
\nabla^2 \phi_i = \frac{\partial^2 \phi_i}{\partial r^2} + \frac{d - 3}{r} \frac{\partial \phi_i}{\partial r}
= \frac{\lambda_*}{6} \phi_a^2 \phi_i
\end{equation}
where the bulk fixed point coupling is given by eq.~(\ref{lambda*}). Considering the ansatz $\<\phi_a(x)\>=c/r$ dictated by conformal symmetry, this admits a solution
\begin{equation}
\<\phi_a(x)\> = 
\begin{cases}
\sqrt{\frac{6}{\lambda_*}} \frac{1}{r}, & a = N \\
0, & a = 1, 2, \cdots, N-1
\label{one-point-break}
\end{cases}
\end{equation}
This is the phase where the $O(N)$ symmetry is broken to $O(N-1)$.

The above analysis can be also carried out in an equivalent way by making a conformal transformation to $H^3 \times S^{d-3}$, as discussed above in the free theory context, see eq.~(\ref{metric-weyl}). After a Weyl rescaling
the bulk action becomes
\begin{equation}\label{action_ads}
S=\int d^{d} x \sqrt{g} \left[\frac{1}{2}\left(\partial \phi_{a}\right)^{2}
+\frac{(d-2)(d-6)}{8} \phi_a^2
+\frac{\lambda_*}{4 !}\left(\phi_{a}^{2}\right)^{2}\right]
\end{equation}
The one point function of a scalar operator is a constant $\<\phi_N\>=c$ in $H^3\times S^{d-3}$ (the Weyl factor cancels the $r$-dependent factor in (\ref{one-point})). Then the $O(N)$ breaking configuration can be simply found by extremizing the potential
\begin{equation}
V(\phi) = \frac{(d-2)(d-6)}{8} \phi_a^2
+\frac{\lambda_*}{4 !}\left(\phi_{a}^{2}\right)^{2}\,.
\end{equation}
This gives
\begin{equation}
\phi_a^2 = \frac{3(6-d)(d-2)}{2 \lambda_*} \implies
\<\phi_a(x)\> = 
\begin{cases}
\sqrt{\frac{3(6-d)(d-2)}{2 \lambda_*}}, & a = N \\
0, & a = \{1, 2, \cdots, N-1\}
\end{cases}
\end{equation}
Transforming back to flat space, we find
\begin{equation}
\<\phi_N(x)\>_{\text{flat}} = \sqrt{\frac{3(6-d)(d-2)}{2 \lambda_*}} \frac{1}{r} = \sqrt{\frac{6}{\lambda_*}} \frac{1}{r}
\end{equation}
to leading order at $d=4-\epsilon$, in agreement with the equation of motion calculation in flat space. 

Expanding the action~(\ref{action_ads}) around this minimum using $\phi_i, i = 1, 2, \cdots, N-1$ and $\phi_N = \sqrt{\frac{3(6-d)(d-2)}{2 \lambda_*}} + \chi$, we find
\begin{equation}
\begin{aligned}
S =& \int d^{d} x \sqrt{g} \left[\frac{1}{2}\left(\partial \phi_i\right)^{2}
+ \frac{1}{2}\left(\partial \chi \right)^{2} 
-\frac{3
   (d-6)^2 (d-2)^2}{32 \lambda }-\frac{1}{4} (d-6) (d-2) \chi ^2 \right. \\
& \left. + \frac{\sqrt{\lambda}}{2} \sqrt{\frac{(6-d)(d-2)}{6}} \left(\chi ^3+\chi  \phi ^2 \right)
+ \frac{1}{24} \lambda  \left(\chi ^2+\phi ^2\right)^2
\right]
\end{aligned}
\end{equation}
Therefore, we have $N-1$ massless scalar fields $\phi_i$ and a massive scalar field $\chi$ with mass $m_\chi^2 = \frac{1}{2} (6-d) (d-2)$.  

According to the familiar mass/dimension relation on $H^3$, $\Delta(\Delta-2)=m^2$, the $N-1$ massless scalars correspond to $N-1$ operators on the defect with $\Delta=2$. These are sometimes referred to as ``tilt" operators: their presence follow from the broken $O(N)$ symmetry and they should have protected dimension. These massless operators correspond to the bottom component in the Kaluza-Klein tower arising from reduction on $S^{d-3}$. The higher states will have $m^2 = \ell(\ell+d-4)$ which leads to $\Delta^{\hat\phi^i}_{\ell} =1+\sqrt{1+\ell^2}$ to leading order in $d=4-\epsilon$. 

From the field $\chi$, we find a tower of states with  $m^2_{\chi} = 2+\ell^2$ (working to leading order, i.e. setting $d=4$), which gives $\Delta^{\hat\chi}_{\ell} = 1+\sqrt{3+\ell^2}$. The $\ell=1$ state has $\Delta^{\hat\chi}_{1} = 3$, and should correspond to the displacement operator with protected scaling dimension. 

From here one may proceed to further analyze the $O(N)$ breaking phase of the planar defect in the $O(N)$ model, but we leave further studies to future work (see~\cite{Krishnan:2023cff} for a recent discussion of the $d=3$ case).

\section{Large $N$}
\label{sec:largeN}
In this section we study the planar defect in the interacting $O(N)$ model at large $N$, keeping $d$ arbitrary. We start with a brief review of the large $N$ treatment of the $O(N)$ model and  a few relevant results for the ``bulk" scaling dimensions, and then we move on to study the planar defect. 

\subsection{Review of large $N$ results}
The large $N$ expansion can be developed by introducing the Hubbard-Stratonovich auxiliary field $\sigma$ as
\begin{equation}\label{S:sigma}
S=\int d^{d} x\left(\frac{1}{2}(\partial \phi_a)^{2}+\frac{1}{\sqrt{N}} \sigma \phi_a^{2} - \frac{6 \sigma^2}{N \lambda_0}\right)
\end{equation}
Integrating out $\sigma$ gives back the original Lagrangian. At the IR fixed point of the $O(N)$ model, the $\sim\sigma^2/\lambda_0$ term can be dropped, and we are left with
\begin{equation}
S=\int d^{d} x\left(\frac{1}{2}(\partial \phi_a)^{2}+\frac{1}{\sqrt{N}} \sigma \phi_a^{2}\right)
\label{N_act}\end{equation}
This action can be used for the $1/N$ expansion of the theory. The free propagator for $\phi_a$ is the same as in eq.~(\ref{prop}) while the ``free" propagator for $\sigma$ (obtained from integrating out $\phi$ at one-loop) is
\begin{equation}
\langle\sigma(x) \sigma(y)\rangle= \frac{C_{\sigma}}{|x-y|^{4-\delta}},
\quad
C_\sigma=\frac{2^{d} \Gamma\left(\frac{d-1}{2}\right) \sin \left(\frac{\pi d}{2}\right)}{\pi^{\frac{3}{2}} \Gamma\left(\frac{d}{2}-2\right)}
\label{C_sigma}\end{equation}
He we have introduced a small shift $\delta$ to the $\sigma$ propagator to regulate the infinities in the divergent conformal graphs (see e.g. \cite{Chai:2021uhv}). At the end of the calculations, we will take $\delta \rightarrow0$ and extract the finite part. The scaling dimensions of $\phi_a$ and $\sigma$ are well-known and given by 
\begin{equation}
\Delta_\phi=\frac{d}{2}-1+\frac{\eta_1}{N}+\mathcal{O}\left(N^{-2}\right),
\quad
\eta_{1} =\frac{2^{d-3}(d-4) \Gamma\left(\frac{d-1}{2}\right) \sin \left(\frac{\pi d}{2}\right)}{\pi^{\frac{3}{2}} \Gamma\left(\frac{d}{2}+1\right)}
\label{Delta_phi0}\end{equation}
\begin{equation}
\Delta_{\sigma}=2+\frac{t}{N}+\mathcal{O}\left(\frac{1}{N^{2}}\right)
\label{Delta_sigma}\end{equation}
where
\begin{equation}
t = \frac{4(d-1)(d-2)}{d-4} \eta_{1}
= \frac{2^d (d-2) \sin \left(\frac{\pi  d}{2}\right) \Gamma \left(\frac{d+1}{2}\right)}{\pi ^{3/2} \Gamma \left(\frac{d}{2}+1\right)}
\label{t}\end{equation}

\subsection{Defect}
\label{largeN}
As $\sigma$ essentially represents the operator $\phi_a^2$ at the IR fixed point, it is natural to define the defect action by
\begin{equation}
S_{d}= h_0 \int d^2 x \,\sigma(\vec{x},0)\,.
\end{equation}
Here we use the same notation $h_0$ for the defect coupling constant, but a priori this coupling is not directly the same as the one used in the $4-\epsilon$ expansion, because the operators $\sigma$ and $\phi_a^2$ have different normalizations. 

In the $O(N)$ invariant DCFT fixed point in $d=4-\epsilon$, we found  that the one-point function of $\phi^2$ has the behavior $\langle \phi^2\rangle \sim \epsilon N^0$ at large $N$, see eq.~(\ref{one-point-phisq}). Note that the $\sigma$ equation of motion from the action~(\ref{S:sigma}) is $\sigma = \frac{\sqrt{N} \lambda}{12} \phi^2$. Since at large $N$ we have $\lambda_* \sim 1/N$, we expect that at the $O(N)$ invariant DCFT fixed point we should have $\langle \sigma \rangle  \sim 1/\sqrt{N}$ in the large $N$ approach. This one-point function is not ``classical" from the point of view of the large $N$ expansion (in the normalizations of eq.~(\ref{S:sigma}), a classical solution of the $\sigma$ effective action behaves as $\sigma \sim \sqrt{N}$), therefore we expect that to match the $O(N)$ invariant fixed point seen in $d=4-\epsilon$, we can simply do $1/N$ perturbation theory around the trivial saddle point $\phi_a = \sigma = 0$.\footnote{This is a saddle point because $\frac{\partial \mathcal{F}}{\partial \sigma}|_{\sigma=0}$ vanishes, since this is proportional to the one-point function of $\phi^2$ in the $\sigma=0$ free theory.} On the other hand, the $O(N)$ breaking fixed point discussed in Section \ref{sym_4} has $\langle \phi^2_a\rangle \sim 1/\lambda_* \sim N$ (see eq.~(\ref{one-point-break})), which translates to $\langle \sigma\rangle \sim \sqrt{N}$, and therefore should correspond to a non-trivial classical saddle point at large $N$. In this paper we will focus on the $O(N)$ invariant phase at large $N$ for generic $d$, and hence we will simply 
do perturbation theory around the trivial configuration. Let us note that the case $d=3$ requires special treatment, as we shall see below, since in that case we expect $\langle \sigma \rangle \sim \sqrt{N}$ even in the $O(N)$ invariant phase, see \cite{Krishnan:2023cff} and also \cite{Giombi:2020rmc} for the large $N$ treatment of the closely related boundary problem (recall that in $d=3$, the planar defect becomes an interface).  

To determine the beta function of the defect coupling, we can now follow a similar procedure as above and calculate the one-point function $\<\sig(0, x)\>_D$, where $x\in \R^{d-2}$ are the coordinates perpendicular to the defect plane.
The leading order diagram is a $\sig$ propagator integrated on the defect plane (diagram $P_0$) while the next diagram $P_1$ is of order $O\left(\frac{h_0^2}{N^{1/2}}\right)$: 
\begin{equation}
\begin{gathered}
\begin{tikzpicture}
\draw[dashed] (0, -1.732) -- (0, 2);
\draw[line width=2] (-1.5, -1.732) -- (1.5, -1.732);
\coordinate[label=below:{$P_0$}] (A) at (0,-1.732);
\end{tikzpicture}
\begin{tikzpicture}
\draw[dashed] (0, 1) -- (0, 2);
\draw (0,0) circle (1);
\draw[dashed] (-1/2, -1.732/2) -- (-1, -1.732);
\draw[dashed] (1/2, -1.732/2) -- (1, -1.732);
\draw[line width=2] (-1.5, -1.732) -- (1.5, -1.732);
\coordinate[label=below:{$P_1$}] (A) at (0,-1.732);
\end{tikzpicture}
\end{gathered}
\label{P_diagrams}
\end{equation}
where we have used dashed line for $\sigma$ propagator, thin solid line for $\phi_a$ propagator, and thick solid line for the defect plane. The calculation of $P_0$ is straightforward:
\begin{equation}
P_0 = - h_0 \int d^2 y \frac{C_\sig}{(x^2 + y^2)^2} = -h_0 C_\sigma \frac{\pi}{x^2} \quad\text{eq.~\ref{I3}}
\end{equation}
To calculate $P_1$, we use the three-point function of $\sigma$
\begin{equation}
\left\langle\sigma\left(x_{1}\right) \sigma\left(x_{2}\right) \sigma\left(x_{3}\right)\right\rangle=\frac{g_{\sigma^{3}}}{\left(\left|x_{12}\right|\left|x_{23}\right|\left|x_{13}\right|\right)^{2-\delta}}
\end{equation}
where~\cite{Chai:2021uhv}
\begin{equation}\label{g_sigma^3}
g_{\sigma^3}=-\frac{1}{\sqrt{N}} \frac{8^{d-1} \sin ^{3}\left(\frac{\pi d}{2}\right) \Gamma\left(3-\frac{d}{2}\right) \Gamma\left(\frac{d-1}{2}\right)^{3}}{\pi^{9 / 2} \Gamma(d-3)}
\end{equation}
The double integral on the defect plane can be calculated by eq.~\ref{I38}:
\begin{equation}
\begin{aligned}
\int d^2 x_2 \int d^2 x_3 
\frac{1}{\left(\left|x_{12}\right|\left|x_{23}\right|\left|x_{13}\right|\right)^{\Delta_{\sigma}}}
= \frac{\pi ^2 \Gamma \left(1-\frac{\Delta_\sigma }{2}\right) \Gamma (\Delta_\sigma -1)^2 \Gamma \left(\frac{3 \Delta_\sigma }{2}-2\right) \left(x_1^2\right){}^{2-\frac{3 \Delta_\sigma }{2}}}{\Gamma (2 (\Delta_\sigma -1)) \Gamma
   \left(\frac{\Delta_\sigma}{2}\right)^2}
\end{aligned}\label{sigma_int}
\end{equation}

So in total we have
\begin{equation}
\begin{aligned}
&\langle\sigma(0, x)\rangle_D 
= -h_0 C_\sigma \frac{\pi}{x^2} \\
& - \frac{h_0^2}{2} 
\frac{1}{\sqrt{N}} \frac{8^{d-1} \sin ^{3}\left(\frac{\pi d}{2}\right) \Gamma\left(3-\frac{d}{2}\right) \Gamma\left(\frac{d-1}{2}\right)^{3}}{\pi^{9 / 2} \Gamma(d-3)}
\frac{\pi ^2 \Gamma \left(1-\frac{\Delta_\sigma }{2}\right) \Gamma (\Delta_\sigma -1)^2 \Gamma \left(\frac{3 \Delta_\sigma }{2}-2\right) \left(x^2\right)^{2-\frac{3 \Delta_\sigma }{2}}}{\Gamma (2 (\Delta_\sigma -1)) \Gamma
   \left(\frac{\Delta_\sigma}{2}\right)^2}\,.
\end{aligned}
\label{sigma_D}\end{equation}
Now we can calculate the renormalized coupling $h$. Substitute
\begin{equation}
\sigma = Z [\sigma], \quad
h_0 = M^{2-\Delta_\sigma } \left(h + \frac{a_{11} h^2}{\delta  \sqrt{N}}\right), \quad
\Delta_{\sigma}=2-\delta
\label{delta_h}\end{equation}
expand in $\frac{1}{\sqrt{N}}$ and $h$, and require the renormalized one-point function $\langle[\sigma](x)\rangle_D$ to be finite as $\delta\to0$ (we set $Z=1$ because $O(1/N)$ correction will not affect the calculation of $a_{11}$). We get
\begin{equation}
a_{11} = \frac{2^d (d-3) \sin ^2\left(\frac{\pi  d}{2}\right) \Gamma \left(2-\frac{d}{2}\right) \Gamma \left(\frac{d-1}{2}\right)}{\pi ^{3/2}}
\label{Z_sigma}\end{equation}
To calculate $\beta_h$, require $\frac{\partial h_0}{\partial\log(M)} = 0$ and bring in the $1/N$ correction $\Delta_{\sigma}=2+\frac{t}{N}$ in eq.~(\ref{delta_h}) (where we now take the regulator $\delta \rightarrow 0$): 
\begin{equation}
\beta_h =\frac{t}{N} h  + \frac{2^d (d-3) h^2 \sin ^2\left(\frac{\pi  d}{2}\right) \Gamma \left(2-\frac{d}{2}\right) \Gamma \left(\frac{d-1}{2}\right)}{\pi ^{3/2} \sqrt{N}}
\label{beta_largeN}\end{equation}

As discussed in previous sections, an alternative way to obtain the beta function is to use the conformal perturbation theory approach. Now the perturbation operator on the defect is $\hat{O} = \hat\sigma$. Let $\varepsilon = 2 - \Delta_\sigma = - t/N$ (eq.~(\ref{Delta_sigma})). The three-point coefficient $g_{\sigma^3}$ is given by eq.~(\ref{g_sigma^3}) and the two-point function coefficient is $C_\sig$ in eq.~(\ref{C_sigma}). Then we get 
\begin{equation}
\begin{aligned}
\beta_{h}
=&-\varepsilon {h}+ \pi \frac{g_{\sigma^3}}{C_\sig} {h}^2+O\left({h}^3\right) \\
=& \frac{t}{N} h + \frac{2^d (d-3) \sin ^2\left(\frac{\pi  d}{2}\right) \Gamma \left(2-\frac{d}{2}\right)
   \Gamma \left(\frac{d-1}{2}\right)}{\pi ^{3/2} \sqrt{N}} h^2 + O(h^3), \\
\end{aligned}
\end{equation}
which is indeed the same as eq.~(\ref{beta_largeN}).

From the beta function we find the fixed point
\begin{equation}
h_* = \frac{2 (d-1)}{\pi  (d-3) d \sqrt{N}}+O(N^{-3/2})
\label{h*}
\end{equation}
As mentioned above, we see that the $d=3$ case requires special care. In this case formally $h_*\rightarrow \infty$ because $P_1$ vanishes, but this just means that the large $N$ and $d\rightarrow 3$ limits do not commute. One may of course set $d=3$ from the start and then develop the $1/N$ expansion. In this case, since the defect becomes an interface, the $O(N)$ invariant defect fixed point is expected to be equivalent to two copies of the so-called ordinary transition in the corresponding boundary problem, as explained in \cite{Krishnan:2023cff}.   Note that the one-point function of $\sigma$ at the fixed point (\ref{h*}) to leading order is
\begin{equation}
\langle\sigma(0, x)\rangle_D
=-\frac{h_* C_{\sigma}\pi }{x^2} = -\frac{2^{d+2} \sin \left(\frac{\pi  d}{2}\right) \Gamma \left(\frac{d+1}{2}\right)}{\pi ^{3/2} (d-3) d \sqrt{N} x^2 \Gamma \left(\frac{d}{2}-2\right)}
\end{equation}
The fact that this diverges for $d=3$ should be related to the fact that the $N$ dependence at large $N$ is enhanced in 3d. Indeed, working directly in $d=3$ one has $\langle \sigma\rangle_D \sim \sqrt{N}$ for the closely related ordinary transition of the BCFT problem. Presumably, to see this from the large $N$ expansion in general $d$, one would need to resum all orders in the $1/N$ expansion, and then take $d=3$.\footnote{For instance, a toy function with the desired properties is of the form $\langle \sigma \rangle \sim \frac{\sqrt{N}}{\sqrt{a+b(d-3)^2N^2}}$, with $a$, $b$ some constants. For $d\neq 3$ this corresponds to a $1/N$ expansion starting at order $1/((d-3)\sqrt{N})$, while if we take $d=3$ first, we get the behavior $\sim \sqrt{N}$.} Alternatively, an additional saddle point for $\sigma$, different from the trivial one we focused on, may appear and be the dominant contribution near $d=3$. 

To compare to the result obtained in the $\epsilon$ expansion, we can consider the normalized one-point function
\begin{equation}
\frac{\langle\sigma(0, x)\rangle_D} {\sqrt{C_\sigma}}
= -\frac{4 \Gamma \left(\frac{d+1}{2}\right) \sqrt{\frac{2^d \sin \left(\frac{\pi  d}{2}\right) \Gamma \left(\frac{d-1}{2}\right)}{\Gamma \left(\frac{d-4}{2}\right)}}}{\pi ^{3/4} (d-3) d \sqrt{N} x^2 \Gamma
   \left(\frac{d-1}{2}\right)}
= -\frac{3 \epsilon }{\sqrt{2} \sqrt{N} x^2}+O(\epsilon^2)
\label{sigma_norm}\end{equation}
where in the second equality we have put in $d=4-\eps$ and expanded in $\epsilon$. Since $\sigma$ stands for the composite field $\phi_a^2$ in the $O(N)$ model with quartic interaction at the IR fixed point, the normalized one-point function of $\sigma$ should be the same as the normalized one-point function of $[\phi_a^2]$. Indeed, we see that it matches eq.~(\ref{phi2_4n}) at the leading order of $\epsilon$ and $\frac{1}{N}$ expansions.


\subsection{Scaling dimensions}
The scaling dimension of $\hat{\phi}_a$ inserted on the defect can be calculated by considering the two-point function $\langle \hat{\phi}_1(y_1) \hat{\phi}_1(y_2)\rangle_d$ where $y_1, y_2 \in \R^2$ are coordinates on the two-dimensional defect plane. The leading corrections to the free propagator are 
\begin{equation}
\begin{gathered}
\begin{tikzpicture}
\draw (1, -1.732) arc(0:180:1 and 1.732);
\draw[dashed] (0, 0) -- (0, -1.732);
\draw[line width=2] (-1.5, -1.732) -- (1.5, -1.732);
\end{tikzpicture}
\begin{tikzpicture}
\draw (1, -1.732) arc(0:180:1 and 1.732);
\draw[dashed] (-0.8, -0.7) -- (0.8, -0.7);
\draw[line width=2] (-1.5, -1.732) -- (1.5, -1.732);
\end{tikzpicture}
\end{gathered}
\label{diags}
\end{equation}
Repeatedly applying eq.~(\ref{I3}) and~(\ref{2prop}), the first diagram is
\begin{equation}
\begin{aligned}
&\frac{2h_0}{\sqrt{N}} \int {d^{d-2} z_1} \int {d^2 z_2} \int {d^2 z_3} \frac{C_\phi^2 C_\sigma}{((z_2-z_3)^2+z_1^2)^{2-\delta} ((z_2-x)^2+z_1^2)^{\frac{d-2}{2}} ((z_2-y)^2+z_1^2)^{\frac{d-2}{2}}} \\
=& \frac{2h_0}{\sqrt{N}} \int {d^{d-2} z_1} \int {d^2 z_2} \frac{C_\phi^2 C_\sigma}{((z_2-x)^2+z_1^2)^{\frac{d-2}{2}} ((z_2-y)^2+z_1^2)^{\frac{d-2}{2}}} 
\frac{\pi  \left(z_1^2\right){}^{\delta -1}}{1 -\delta} \\
=& \frac{2h_0}{\sqrt{N}} \int {d^{d-2} z_1} \int_0^1 ds {C_\phi^2 C_\sigma}
\frac{\pi  \left(z_1^2\right){}^{\delta -1}}{1 -\delta} 
\frac{\pi  (1-s)^{\frac{d-2}{2}-1} s^{\frac{d-2}{2}-1} \Gamma (d-3) \left((1-s) s y^2+z_1^2\right){}^{3-d}}{\Gamma \left(\frac{d-2}{2}\right)^2} \\
=& \frac{2h_0}{\sqrt{N}} \int_0^1 ds {C_\phi^2 C_\sigma}
\frac{\pi}{1 -\delta} 
\frac{\pi  (1-s)^{\frac{d-2}{2}-1} s^{\frac{d-2}{2}-1} \Gamma (d-3)}{\Gamma \left(\frac{d-2}{2}\right)^2} \\
& \times \frac{\pi ^{\frac{d-2}{2}} \Gamma \left(\frac{d}{2}+\delta -2\right) \Gamma \left(\frac{1}{2} (d-2 (\delta +1))\right) \left(-\left((s-1) s y^2\right)\right)^{-\frac{d}{2}+\delta +1}}{\Gamma
   \left(\frac{d}{2}-1\right) \Gamma (d-3)} \\
=& -\frac{\pi ^{-d/2} h_0 2^{d-2 (\delta +1)} \sin \left(\frac{\pi  d}{2}\right) \Gamma \left(\frac{d-1}{2}\right) \Gamma (\delta -1) \Gamma \left(\frac{d}{2}+\delta -2\right) \Gamma \left(\frac{1}{2} (d-2
   (\delta +1))\right) \left(y^2\right)^{-\frac{d}{2}+\delta +1}}{\sqrt{N} \Gamma \left(\frac{d}{2}-1\right) \Gamma \left(\frac{d-4}{2}\right) \Gamma \left(\delta +\frac{1}{2}\right)}
\end{aligned}
\label{Delta_phi}
\end{equation}
where $y = y_1 - y_2$. Since the anomalous dimension $\gamma_{\hat{\phi}}$ will manifest as a correction in the exponent of the free propagator
\begin{equation}
\frac{C_\phi}{(y^2)^{\frac{d-2}{2}+\gamma_{\hat{\phi}}}}
= C_\phi \left(y^2\right)^{1-\frac{d}{2}}-\gamma_{\hat{\phi}}  C_\phi \left(y^2\right)^{1-\frac{d}{2}} \log (y^2)+O\left(\gamma^2_{{\hat{\phi}}}\right)
\end{equation}
we can expand eq.~\ref{Delta_phi} in $\delta$ and extract the coefficient of $- C_\phi \left(y^2\right)^{1-\frac{d}{2}} \log (y^2)$, which gives us
\begin{equation}
\gamma_{\hat{\phi}}|_{\rm diag.~1} = -\frac{2^d \pi ^{\frac{1}{2} (-d-1)+\frac{d}{2}} h_0 \sin \left(\frac{\pi  d}{2}\right) \Gamma \left(\frac{d-1}{2}\right)}{\sqrt{N} \Gamma \left(\frac{d-2}{2}\right)}
\end{equation}
The second diagram is exactly the leading correction to the free propagator in the bulk theory (without defect), so its contribution is $\gamma_{\hat{\phi}}|_{\rm diag.~2} = \frac{\eta_1}{N}$ as in eq.~(\ref{Delta_phi0}). In total, after replacing $h_0$ by $h$ which is consistent to this order, we have
\begin{equation}
\gamma_{\hat{\phi}} = -\frac{2^d \sin \left(\frac{\pi  d}{2}\right) \Gamma \left(\frac{d-1}{2}\right)}{\sqrt{N} \pi^{\frac{1}{2}}\Gamma \left(\frac{d-2}{2}\right)} h
+ \frac{2^{d-3}(d-4) \Gamma\left(\frac{d-1}{2}\right) \sin \left(\frac{\pi d}{2}\right)}{N \pi^{\frac{3}{2}} \Gamma\left(\frac{d}{2}+1\right)}
\label{gamma_phi_largeN}
\end{equation}
At the fixed point $h=h_*$ in (\ref{delta_h}), we then have
\begin{equation}
\Delta_{{\hat{\phi}}} =\frac{d}{2}-1+\frac{\sin \left(\frac{\pi  d}{2}\right) \left(\frac{\sqrt{\pi } (d-4) \Gamma (d+1)}{d-1}-\frac{2^{d+4} \Gamma \left(\frac{d+1}{2}\right)}{(d-3) d \Gamma \left(\frac{d}{2}-1\right)}\right)}{4N \pi ^{3/2}}+O(1/N^2)\,.
\end{equation}
Setting $d=4-\epsilon$ and expanding in small $\epsilon$, this gives
\begin{equation}
\Delta_{{\hat{\phi}}} = 1-\frac{\epsilon}{2}+\frac{6\epsilon}{N}+O(1/N^2)\,. 
\end{equation}
This matches eq.~(\ref{Deltaphi}) to to order $1/N$, which provides a non-trivial consistency check. 

 The dimension of $\hat{\sigma}$ inserted on the defect  can be computed from the derivative of the beta function:
\begin{equation}
\Delta_{\hat{\sigma}} = 2 + \left.\frac{\partial \beta_h}{\partial h}\right\vert_{h=h_*} 
= 2 - \frac{t}{N}
\label{Delta_phi^2_N}
\end{equation}
In the large $N$ theory, $\sigma$ represents $\phi_a^2$ at the IR fixed point so $\Delta_{\hat{\sigma}}$ should match $\Delta_{{\hat{\phi}}^2}$ (eq.~\ref{delta_phi2_4}) computed in $d=4-\eps$. Indeed, setting $d=4-\eps$ and expanding in $\epsilon$ we get 
\begin{equation}
\Delta_{\hat{\sigma}}  = 2 + \frac{6 \epsilon }{N}-\frac{13 \epsilon ^2}{2 N}+\ldots 
\end{equation}
This indeed matches eq.~(\ref{delta_phi2_4}) to $O\left(\frac{\epsilon}{N}\right)$. 
Interestingly, $\hat{\phi}$ and $\hat{\sigma}$ have the same anomalous dimension to $O\left(\frac{\epsilon}{N}\right)$.

\section{Cubic model in $d=6-\epsilon$}
\label{sec:6-eps}
Consider the following Euclidean action: 
\begin{equation}
S=\int d^d x \left[\frac{1}{2}\left(\partial_\mu \phi_a\right)^2+\frac{1}{2}\left(\partial_\mu \sigma\right)^2+\frac{g_1}{2} \sigma \phi^2_a +\frac{g_2}{6} \sigma^3 \right]
\end{equation}
This model has an IR fixed point in $d=6-\eps$ which is the same as the UV fixed point of the quartic theory in $d>4$ \cite{Fei:2014yja}. This fixed point is unstable nonperturbatively due to instanton corrections \cite{Giombi:2019upv}, but here our main interest is to use the cubic model description within perturbation theory as a further consistency check of the large $N$ results. 

The surface defect action in this theory has the same form as in large $N$ theory 
\begin{equation}
S_D = h_0 \int d^2 x\, \sigma(x, 0)\,.
\end{equation}
This kind of surface defect in the cubic scalar field theory was also previously studied in \cite{Rodriguez-Gomez:2022gbz, Rodriguez-Gomez:2022gif, CarrenoBolla:2023sos}, in a certain semiclassical double-scaling limit. 

Consider $\langle\sigma(0, x)\rangle_D$ where $x\in\R^{d-2}$ are the coordinates perpendicular to the defect. We assume the fixed point renormalized coupling $h_*$ is the same order as $g_1^*, g_2^*$, which we will confirm later. Using the same notation as in eq.~(\ref{P_diagrams}), the leading diagrams in $\langle\sigma(0, x)\rangle_D$ are:
\begin{equation}
\begin{gathered}
\begin{tikzpicture}
\draw[dashed] (0, -1.732) -- (0, 2);
\draw[line width=2] (-1.5, -1.732) -- (1.5, -1.732);
\coordinate[label=below:{$Q_0$}] (A) at (0,-1.732);
\end{tikzpicture}
\begin{tikzpicture}
\draw[dashed] (0, 1) -- (0, 2);
\draw (0,0) circle (1);
\draw[dashed] (0, -1.732) -- (0, -1);
\draw[line width=2] (-1.5, -1.732) -- (1.5, -1.732);
\coordinate[label=below:{$Q_1$}] (A) at (0,-1.732);
\end{tikzpicture}
\begin{tikzpicture}
\draw[dashed] (0, 1) -- (0, 2);
\draw[dashed] (0,0) circle (1);
\draw[dashed] (0, -1.732) -- (0, -1);
\draw[line width=2] (-1.5, -1.732) -- (1.5, -1.732);
\coordinate[label=below:{$Q_2$}] (A) at (0,-1.732);
\end{tikzpicture}
\begin{tikzpicture}
\draw[dashed] (0, 0) -- (0, 2);
\draw[dashed] (0, 0) -- (-1, -1.732);
\draw[dashed] (0, 0) -- (1, -1.732);
\draw[line width=2] (-1.5, -1.732) -- (1.5, -1.732);
\coordinate[label=below:{$Q_3$}] (A) at (0,-1.732);
\end{tikzpicture}
\end{gathered}
\end{equation}


To proceed, it is easiest to perform the calculation in momentum space. We consider $\<\sig(0, p)\>_D$ where $p\in\R^{d-2}$ is the momentum perpendicular to the defect plane. $Q_1 + Q_2$ can be calculated by eq.~(\ref{I1})
\begin{equation}
-\frac{h_0 (g_2^2 + N g_1^2)}{2} \int\frac{d^d q} {(2\pi)^d}\frac{1}{p^4 q^2 (p-q)^2}
= \frac{h_0 (g_2^2 + N g_1^2)}{2} \frac{2^{3-2 d} \pi ^{\frac{3-d}{2}} p^{d-8} \csc \left(\frac{\pi  d}{2}\right)}{\Gamma \left(\frac{d-1}{2}\right)}
\end{equation}
while $Q_3$ is
\begin{equation}
\begin{aligned}
- \frac{h_0^2 g_2}{2} \int \frac{d^{d-2} q} {(2\pi)^{d-2}}\frac{1}{p^2 q^2 (p-q)^2} = - {h_0^2 g_2} \frac{4^{3-d} \pi ^{\frac{5-d}{2}} p^{d-8} \csc \left(\frac{\pi  d}{2}\right)}{\Gamma \left(\frac{d-3}{2}\right)}
\end{aligned}
\end{equation}
In total,
\begin{equation}
\begin{aligned}
\langle\sigma(0,p)\rangle_D 
&= \frac{-h_0}{p^2} - {h_0^2 g_2} \frac{4^{3-d} \pi ^{\frac{5-d}{2}} p^{d-8} \csc \left(\frac{\pi  d}{2}\right)}{\Gamma \left(\frac{d-3}{2}\right)}
+ \frac{h_0}{(4 \pi)^3} \frac{N g_1^2+g_2^2}{12} \frac{\Gamma(3-d / 2)}{(p^2)^{4-d / 2}}
\end{aligned}
\end{equation}
The $Z$ factor for the bulk operator $\sigma$ can be calculated by renormalizing the 2-point function (in the absence of the defect). This has been calculated in~\cite{Fei:2014yja}:
\begin{equation}
\langle\sigma(p) \sigma(-p)\rangle = \frac{1}{p^2}
- \frac{g_2^2 + N g_1^2}{2} \frac{2^{3-2 d} \pi ^{\frac{3-d}{2}} p^{d-8} \csc \left(\frac{\pi  d}{2}\right)}{\Gamma \left(\frac{d-1}{2}\right)}
\end{equation}
Let $\sigma(p) = Z \times [\sigma](p)$, and require the renormalized field's propagator $\langle[\sigma](p) [\sigma](-p)\rangle$ to be finite as $\eps \rightarrow 0$, we get
\begin{equation}
Z = 1 - \frac{1}{(4 \pi)^3} \frac{N g_1^2+g_2^2}{12 \epsilon} 
\end{equation}
Substitute in
\begin{equation}
h_0 = M^{\frac{\epsilon}{2}} \left(h + a_1 \frac{h^2 {g}_2}{\epsilon} + a_2 \frac{h N {g}^2_1}{\epsilon} + a_3 \frac{h {g}^2_2}{\epsilon}\right)
\end{equation}
and require $\langle[\sigma](0,p)\rangle_D$ to be finite as $\eps \rightarrow 0$, we arrive at:
\begin{equation}
h_0 = M^{\frac{\epsilon}{2}} \left(h - \frac{1}{16\pi^2} \frac{h^2 g_2}{\epsilon} + \frac{1}{768\pi^3} \frac{h N g_1^2}{\epsilon} + \frac{1}{768\pi^3} \frac{h g^2_2}{\epsilon}\right)
\end{equation}
Now we require the bare coupling $h_0$ to be independent of energy scale $M$ and use $\beta_1=-\frac{\epsilon}{2} g_1$, $\beta_2=-\frac{\epsilon}{2} g_2$ to this order.
We then find the beta function
\begin{equation}
\beta_h = -\frac{h \epsilon }{2} -\frac{g_2 h^2}{16 \pi ^2}+\frac{g_1^2 h N}{768 \pi ^3}+\frac{g_2^2 h}{768 \pi ^3}
\label{beta_6-eps}
\end{equation}
At large $N$ the fixed point is:
\begin{equation}
h_* =  \frac{(g_1^*)^2 N+(g_2^*)^2-384 \pi ^3 \epsilon}{48 \pi g^*_2}
= \frac{20 \sqrt{\frac{2 \pi }{3}} \sqrt{\epsilon }}{3 \sqrt{N}}
\label{h_fixed}\end{equation}
where we have put in the fixed point couplings $g_1^*=\sqrt{\frac{6 \epsilon(4 \pi)^3}{N}} (1+22/N+\ldots),\ g_2^*=6 \sqrt{\frac{6 \epsilon(4 \pi)^3}{N}} (1+162/N+\ldots )$~\cite{Fei:2014yja}.
At this fixed point, we have
\begin{equation}
\langle[\sigma](0,p)\rangle_D
= -\frac{20 \sqrt{\frac{2 \pi }{3}} \sqrt{\epsilon }}{3 \sqrt{N} p^2}
\end{equation}
Fourier transforming this to position space using eq.~(\ref{fourier}) gives us:
\begin{equation}
\langle[\sigma](0,x)\rangle_D
= -\frac{5 \sqrt{\frac{2}{3}} \sqrt{\epsilon }}{3 \pi ^{3/2} \sqrt{N} x^2}\,.
\end{equation}
Since to leading order
\be
\langle[\sigma] (x) [\sigma] (y) \rangle = \frac{1}{4 \pi^3 |x-y|^4}
\ee
The normalized one-point function is
\begin{equation}
\frac{\langle[\sigma](0,x)\rangle_D} {\sqrt{1/(4 \pi^3)}}
= -\frac{10 \sqrt{\frac{2}{3}} \sqrt{\epsilon }}{3 \sqrt{N} x^2}
\end{equation}
This agrees with the large $N$ calculation in eq.~(\ref{sigma_norm}) expanded in $d=6-\epsilon$:
\begin{equation}
\frac{\langle\sigma(x)\rangle_D} {\sqrt{\langle\sigma\sigma\rangle}}
= -\frac{4 \Gamma \left(\frac{d+1}{2}\right) \sqrt{\frac{2^d \sin \left(\frac{\pi  d}{2}\right) \Gamma \left(\frac{d-1}{2}\right)}{\Gamma \left(\frac{d-4}{2}\right)}}}{\pi ^{3/4} (d-3) d \sqrt{N} x^2 \Gamma
   \left(\frac{d-1}{2}\right)}
= -\frac{10 \sqrt{\frac{2}{3}} \sqrt{\epsilon }}{3 \sqrt{N} x^2}\,.
\end{equation}

\paragraph{Conformal perturbation theory approach.}
As in previous sections, as a further check let us also use conformal perturbation theory to calculate the beta function. The scaling dimension of $\sig$ at $h=0$ is
\begin{equation}
\Delta_\sigma
=2-\frac{\epsilon}{2}+\frac{1}{(4 \pi)^3} \frac{N\left(g_1^*\right)^2+\left(g_2^*\right)^2}{12}
\end{equation}
So $\varepsilon = 2 - \Delta_\sigma = \frac{\epsilon}{2} -\frac{1}{(4 \pi)^3} \frac{N\left(g_1^*\right)^2+\left(g_2^*\right)^2}{12}$. The three-point function coefficient can be calculated by (see eq.~(\ref{triangle}))
\begin{equation}
\begin{aligned}
\<\sig(x_1) \sig(x_2) \sig(x_3)\>
=& {-g_2} \int d^6 x \frac{C_\phi^3}{|x-x_1|^4 |x-x_2|^4 |x-x_3|^4} \\
=& \frac{-g_2 C_\phi^3 \pi^3}{|x-x_1|^2 |x-x_2|^2 |x-x_3|^2}
\end{aligned}
\end{equation}
Therefore, when the bulk theory is critical,
\begin{equation}
\begin{aligned}
\beta_{h}
=&-\delta {h}+ \pi \frac{-g_2 C_\phi^3 \pi^3}{C_\phi} {h}^2+O\left({h}^3\right) \\
=& -\frac{\epsilon}{2} h +\frac{1}{(4 \pi)^3} \frac{N\left(g_1^*\right)^2+\left(g_2^*\right)^2}{12} h
- \frac{g_2^*}{16 \pi^2} h^2 + O(h^3) \\
\end{aligned}
\end{equation}
where $C_\phi = \frac{1}{4} \pi ^{-d/2} \Gamma \left(\frac{d-2}{2}\right) = \frac{1}{4 \pi^3}$ is evaluated at $d = 6$. This agrees with eq.~(\ref{beta_6-eps}).

\subsection{Scaling dimensions on the defect}
The scaling dimension of $\hat{\sig}$ on the defect at the fixed point is
\begin{equation}
\Delta_{\hat{\sigma}} = 2 + \left.\frac{\partial \beta_h}{\partial h}\right.\vert_{h=h_*, {g}_2 = g_2^*, {g}_1 = g_1^*} 
\end{equation}
If we plug in the explicit fixed point couplings and expand at large $N$, we obtain $\Delta_{\hat{\sigma}} =2- \frac{40 \epsilon }{N}$. It is straightforward to check 
that this matches the large $N$ result (\ref{Delta_phi^2_N}), (\ref{t}) expanded in $d=6-\epsilon$. 


To determine the dimension of $\hat{\phi}_a$ on the defect, let us consider as before $\langle \hat{\phi}_1(y_1) \hat{\phi}_1(y_2)\rangle_D$ where $y_1, y_2 \in \R^2$ are coordinates on the two-dimensional defect plane. The two leading order diagrams are the same as in large $N$ theory (eq.~\ref{diags}):
\begin{equation}
\begin{gathered}
\begin{tikzpicture}
\draw (1, -1.732) arc(0:180:1 and 1.732);
\draw[dashed] (0, 0) -- (0, -1.732);
\draw[line width=2] (-1.5, -1.732) -- (1.5, -1.732);
\end{tikzpicture}
\begin{tikzpicture}
\draw (1, -1.732) arc(0:180:1 and 1.732);
\draw[dashed] (-0.8, -0.7) -- (0.8, -0.7);
\draw[line width=2] (-1.5, -1.732) -- (1.5, -1.732);
\end{tikzpicture}
\end{gathered}
\end{equation}
Define $y = y_1 - y_2$. The first diagram is
\begin{equation}
\begin{aligned}
& {g_1 h_0} \int {d^{d-2} z_1} \int {d^2 z_2} \int {d^2 z_3} 
\frac{C_\phi^3} {((z_2-z_3)^2+z_1^2)^{\frac{d-2}{2}} ((z_2-y)^2+z_1^2)^{\frac{d-2}{2}} (z_2^2+z_1^2)^{\frac{d-2}{2}}} \quad\text{eq.~\ref{I3}}\\
=& {g_1 h_0} \int {d^{d-2} z_1} \int {d^2 z_2} 
\frac{C_\phi^3} {((z_2-y)^2+z_1^2)^{\frac{d-2}{2}} (z_2^2+z_1^2)^{\frac{d-2}{2}}}
\frac{2 \pi  z_1^{4-d}}{d-4}
\quad\text{eq.~\ref{2prop}}\\
=& {g_1 h_0} \int {d^{d-2} z_1} 
{C_\phi^3} 
\frac{2 \pi  z_1^{4-d}}{d-4}
\int_0^1 ds
\frac{\pi  (-((s-1) s))^{\frac{d-4}{2}} \Gamma (d-3) \left(z_1^2-(s-1) s y^2\right)^{3-d}}{\Gamma \left(\frac{d-2}{2}\right)^2}
\quad\text{eq.~\ref{I3}}\\
=& {g_1 h_0} 
{C_\phi^3} 
\frac{2 \pi}{d-4}
\int_0^1 ds
\frac{\pi  (-((s-1) s))^{\frac{d-4}{2}} \Gamma (d-3)}{\Gamma \left(\frac{d-2}{2}\right)^2}
\frac{\pi ^{\frac{d-2}{2}} (s-1)^4 s^4 y^8 \left(-\left((s-1) s y^2\right)\right)^{-d}}{(d-4) \Gamma \left(\frac{d}{2}-1\right)}
\\
=& {g_1 h_0}
\frac{\pi ^{1-d} y^8 \left(y^2\right)^{-d} \Gamma \left(3-\frac{d}{2}\right)^2 \Gamma (d-3)}{32 (d-4)^2 \Gamma (6-d)}
\end{aligned}
\end{equation}
The second diagram has been calculated in momentum space in \cite{Fei:2014yja}:
\begin{equation}
- \frac{p^2}{(4 \pi)^3 p^4} \frac{g_1^2}{6} \frac{\Gamma(3-d / 2)}{\left(p^2\right)^{3-d / 2}}
\end{equation}
It is Fourier transformed as (eq.~\ref{fourier}):
\begin{equation}
- \frac{1}{(4 \pi)^3} \frac{g_1^2}{6} \frac{\Gamma(3-d / 2)}{4 \pi^3 y^4}
\end{equation}
So in total, 
\begin{equation}
\langle\hat{\phi}_1(0)\hat{\phi}_1(y)\rangle_D = \frac{C_\phi}{y^{d-2}} 
+ {g_1 h_0}
\frac{\pi ^{1-d} y^8 \left(y^2\right)^{-d} \Gamma \left(3-\frac{d}{2}\right)^2 \Gamma (d-3)}{32 (d-4)^2 \Gamma (6-d)}
- \frac{1}{(4 \pi)^3} \frac{g_1^2}{6} \frac{\Gamma(3-d / 2)}{4 \pi^3 y^4}
\end{equation}

Now defining
\begin{equation}
\hat{\phi}(y) = Z [\hat{\phi}](y), \quad Z = 1 + a \frac{g_1 h}{\epsilon} + b \frac{g_1^2}{\epsilon}
\end{equation}
and requiring $\langle[\hat{\phi}]_1(0)[\hat{\phi}]_1(y)\rangle_D$ to be finite as $\epsilon\rightarrow 0$, we get
\begin{equation}
Z = 1 + \frac{1}{8 \pi^2} \frac{g_1 h}{\epsilon} - \frac{1}{384 \pi^3} \frac{g_1^2}{\epsilon}
\end{equation}
Therefore the anomalous dimension is
\begin{equation}
\gamma_{\hat{\phi}} = \frac{\partial\log(Z)}{\partial\log(M)}
= \beta_h \frac{\partial\log(Z)}{\partial h} + \beta_{g_1} \frac{\partial\log(Z)}{\partial g_1} 
= \frac{g_1^2}{384 \pi ^3}-\frac{g_1 h}{8 \pi ^2}
\end{equation}
At the fixed point (\ref{h_fixed}), and keeping the leading order at large $N$, this gives
\begin{equation}
\gamma_{\hat{\phi}} 
= -\frac{37 \epsilon}{3 N}+\ldots 
\end{equation}
This agrees with eq.~(\ref{gamma_phi_largeN}) obtained from the large $N$ expansion. 

It is also straightforward to construct the phase of the model that breaks the $O(N)$ symmetry, following the same steps as in Section~\ref{sym_4} (either solving the equations of motion in flat space, or mapping the problem to $H^3\times S^{d-3}$). We will omit the details of the calculation here, but let us mention that, similarly to the boundary case discussed in \cite{Giombi:2020rmc}, the corresponding phase appears to be non-unitary (the classical value of $\phi_N$ turns out to be complex).

\section{Defect Free Energy}
\label{sec:F}
In this section we study the defect free energy in the case where the defect is a two-dimensional sphere. The defect free energy is defined by
\begin{equation}
\cF = -\log \frac{Z_{\mathcal{D}}}{Z_{\mathrm{CFT}}}
\end{equation}
where $Z_{\cD}$ is the partition function of the CFT in the presence of a spherical defect, and $Z_{\mathrm{CFT}}$ is the partition function of the CFT without the defect. On general grounds, for a sphere of radius $R$, the defect free energy takes the form
\begin{equation}
\cF = a_1 +a_2 (MR)^2 -\frac{b}{3} \log(MR)
\label{Fvsb}
\end{equation}
where $M$ is a renormalization scale. The coefficients $a_1$ and $a_2$ are non-universal (scheme dependent), while $b$ is a defect Weyl anomaly coefficient which is scheme independent. It is a direct analog of the central charge $c$ in a 2d CFT. It was proved in \cite{Jensen:2015swa} (see also \cite{Shachar:2022fqk} for an alternative proof) that the $b$ coefficient decreases under defect RG flow 
\begin{equation}
b_{\rm UV} > b_{\rm IR}\,.
\label{b-theorem}
\end{equation}

Below we will compute the exact $b$ coefficient at the IR fixed point of the defect RG flow in the free theory discussed in Section \ref{sec:free}, and then compute $b$ perturbatively in the case of the interacting $O(N)$ model in $d=4-\epsilon$. In both cases the results are consistent with (\ref{b-theorem}), as expected.


\subsection{Free theory}
Let us start with a perturbative computation of the defect free energy in the free theory. Up to order $h^3$, there are two diagrams that contribute
\begin{equation}
\cF = -F_0 -F_1 +O(h_0^4)
\end{equation}
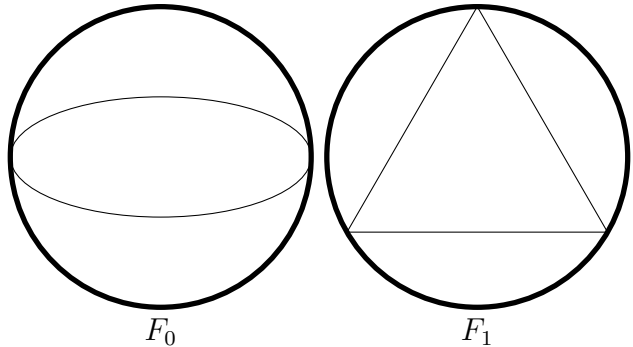
\begin{center}
\begin{tikzpicture}
\coordinate[label=below:{$F_0$}] (A) at (0, -2);
\draw[line width=2] (0, 0) circle (2);
\draw (0, 0) ellipse (2 and 0.8);
\end{tikzpicture}
\begin{tikzpicture}
\coordinate[label=below:{$F_1$}] (A) at (0, -2);
\draw[line width=2] (0, 0) circle (2);
\draw (0, 2) -- (1.732, -1);
\draw (0, 2) -- (-1.732, -1);
\draw (1.732, -1) -- (-1.732, -1);
\end{tikzpicture}
\end{center}
where we have used a big thick circle to represent the spherical defect and the thin lines are free propagators of $\phi_a$. The first diagram is given by
\begin{equation}
\begin{aligned}
F_0
=& \int_D d^2 x \int_D d^2 y \ h_0^2 \frac{N C_\phi^2}{|x-y|^{2(d-2)}} \\
=& 4\pi h_0^2 R^4 \int_0^\pi d\theta \sin(\theta) \int_0^{2\pi} d\phi \frac{N C_\phi^2}{|2R\sin(\frac{\theta}{2})|^{2(d-2)}} \\
=& -\frac{2^{4-2 d} \pi ^{2-d} h_0^2 N R^{8-2 d} \Gamma \left(\frac{d-2}{2}\right)^2}{d-3} 
\label{F0}
\end{aligned}
\end{equation}
where the integral has been evaluated by analytic continuation in $d$. 
Note that this is finite when $d=4$. Using the integral~\cite{Giombi:2014xxa}
\begin{equation}
\int d^d x d^d y d^d z \sqrt{g_x} \sqrt{g_y} \sqrt{g_z} \frac{1}{[s(x, y) s(y, z) s(z, x)]^{\Delta}}=R^{3(d-\Delta)} \frac{8 \pi^{\frac{3(1+d)}{2}} \Gamma\left(d-\frac{3 \Delta}{2}\right)}{\Gamma(d) \Gamma\left(\frac{1+d-\Delta}{2}\right)^3}
\end{equation}
the diagram $F_1$ can be calculated as
\begin{equation}
\begin{aligned}
F_1 =& \int_D d^2 x \int_D d^2 y \int_D d^2 z \ (-h_0^3) \frac{8}{3!} \frac{N C_\phi^3}{|x-y|^{d-2} |z-y|^{d-2} |x-z|^{d-2}} \\
=& -\frac{1}{3} 2^{11-3 d} \pi ^{3-\frac{3 d}{2}} h_0^3 N R^{-3 (d-4)} \cos ^3\left(\frac{\pi  d}{2}\right) \Gamma \left(5-\frac{3 d}{2}\right) \Gamma (d-3)^3\,.
\label{F1}
\end{aligned}
\end{equation}
Putting these two contributions together, replacing the bare coupling $h_0$ by the renormalized one $h_0= M^{\epsilon} h/(1-h/(\pi \epsilon))$, expanding in $\epsilon$ to the relevant order, and focusing on the $\log(MR)$ dependent term, we find 
\begin{equation}
\cF = N \left(\frac{\epsilon h^2}{8\pi^2}-\frac{h^3}{12\pi^3}\right) \log(MR)+\ldots 
\label{cFlogR-free}
\end{equation}
From this and eq.~(\ref{Fvsb}), we can read off the $b$ coefficient at the IR fixed point 
\begin{equation}
b_{\rm IR} = N\left(-\frac{3\epsilon h_*^2}{8\pi^2}+\frac{h_*^3}{4\pi^3}\right) 
= -\frac{N\epsilon^3}{8}\,.
\label{bpert}
\end{equation}
The perturbative evaluation of $b_{\rm IR}$ in the free theory was also recently given in \cite{Shachar:2022fqk}. 

Even though it is not obvious from the above perturbative calculation, the result (\ref{bpert}) is in fact exact (similarly to the way in which the beta function (\ref{free_beta}) is exact). This can be seen by computing the exact free energy using functional determinant techniques, and exploiting the analogy of this defect RG flow to the general problem of double-trace deformations in CFT. To compute the exact free energy, it is convenient to first rephrase the theory (\ref{free_action}) as a non-local 2d theory, by integrating out the bulk degrees of freedom. It is convenient to dimensionally continue the defect from 2 to $p$ dimensions, so that $p=2-\varepsilon$ can serve as a regulator, and we keep the bulk dimension $d$ arbitrary. Let us first work out the 2d action in the case of flat space. The momentum space propagator, restricted to the $p$-dimensional subspace, is given by 
\begin{equation}
\langle \phi_a(-k)\phi_b(k)\rangle = \delta_{ab} \int \frac{d^{d-p}q}{(2\pi)^{d-p}} \frac{1}{k^2+q^2} =\delta_{ab} \frac{2^{p-d} \pi ^{\frac{p-d}{2}} \Gamma \left(\frac{1}{2} (-d+p+2)\right)}{k^{2+p-d}}\,.
\end{equation}
Then we can write the action on the planar $p$-dimensional defect as 
\begin{equation}
S=S_0 + S_h
\end{equation}
with 
\begin{equation}
S_h = h_0 \int d^p x \phi_a\phi_a\,.
\end{equation}
and
\begin{equation}
\begin{aligned}
S_0 =& \int \frac{d^p k}{(2\pi)^p} \phi_a(-k) \frac{1}{2} \frac{k^{2+p-d}}{2^{p-d} \pi ^{\frac{p-d}{2}} \Gamma \left(\frac{1}{2} (-d+p+2)\right)} \phi_a(k) \\
=& \int d^p x \int d^p y \phi_a(x) \frac{1}{2} \left[2 C
|x-y|^{d-2p-2}\right] \phi_a(y)\,,
\end{aligned}
\end{equation}
where we defined
\begin{equation}
C =  (d-p-2) \pi ^{\frac{1}{2} (d-2 (p+1))} 
\sin \left(\frac{1}{2} \pi  (d-p)\right) \Gamma
   \left(-\frac{d}{2}+p+1\right)\,.
\end{equation}

Now mapping this to the sphere, we have $S_h = R^p \int d^p x \sqrt{g_x} h_0 \phi_a\phi_a$ and 
\begin{equation}
\begin{aligned}
S_0 =& R^{d-2} \int d^p x \sqrt{g_x} \int d^p y \sqrt{g_y} \phi_a(x)\frac{1}{2} \left[2C
s(x, y)^{d-2p-2}
\right] \phi_a(y)
\end{aligned}
\end{equation}
where the chordal distance and metric on the unit sphere are
\begin{equation}
s(x, y)=\frac{2 |x-y|}{\left(1+x^2\right)^{1 / 2}\left(1+y^2\right)^{1 / 2}}, 
\quad
g_{\mu\nu} = \frac{4}{\left(1+x^2\right)^2} \delta_{\mu\nu}\,.
\end{equation}
The calculation of the defect free energy then boils down to computing the 
determinant of the non-local operator 
\begin{equation}
R^{d-p-2} \int d^p y \sqrt{g_y} \frac{2 C}{s(x,y)^{2p+2-d}} \phi(y) + 2h_0 \phi(x) = \lambda \phi(x)
\end{equation}
Using the following decomposition in spherical harmonics (see e.g. \cite{Giombi:2019upv})
\begin{equation}
\frac{1}{s\left(x, y\right)^{2 \alpha}}=\sum_{n,\vec{m}}^{\infty} k_n(\alpha) Y_{n, \vec{m}}^*\left(x\right) Y_{n, \vec{m}}\left(y\right), \quad k_n(\alpha)
=\pi^{\frac{p}{2}} 2^{p-2 \alpha} \frac{\Gamma\left(\frac{p}{2}-\alpha\right) \Gamma(n+\alpha)}{\Gamma(\alpha) \Gamma(p+n-\alpha)}
\end{equation}
the eigenvalues of the non-local operator are given by
\begin{equation}
\lambda_n = 2 C R^{d-p-2} k_n\left(p+1-\frac{d}{2}\right) + 2h_0
\end{equation}
Let $D_n^p=\frac{(2 n+p-1) \Gamma(n+p-1)}{n! \Gamma(p)}$ the degeneracy factor for the spherical harmonics. Then the exact defect free energy is given by
\begin{equation}
\begin{aligned}
\cF = & -\log \frac{Z^{(\mathrm{DCFT})}\left[\mathbb{S}^d\right]}{Z^{(\mathrm{CFT})}\left[\mathbb{S}^d\right]} \\
=& \frac{N}{2} \sum_n D_n^p \log \frac{\lambda_n^{h_0}}{\lambda_n^{h_0=0}} 
= \frac{N}{2} \sum_n D_n^p \log\left(1+\frac{h_0 R^{p+2-d}}{C k_n}\right) \\
=& \frac{N}{2} \sum_n D_n^p \log\left(1 + \frac{2^{-d+p+1} \pi ^{\frac{1}{2} (-d+p+2)} \Gamma \left(\frac{d}{2}+n-1\right) \csc
   \left(\frac{1}{2} \pi  (d-p)\right)}{\Gamma \left(\frac{d-p}{2}\right) \Gamma
   \left(-\frac{d}{2}+n+p+1\right)} h_0 R^{p+2-d}\right)\,.
\end{aligned}
\end{equation}
In the IR limit, the second term inside the logarithm dominates. Using also the fact that in dimensional continuation one has the identity $\sum_{n=0}^{\infty} D_n^p = 0$ \cite{Diaz:2007an}, we can simplify the free energy in the IR limit to 
\begin{equation}
\cF_{\rm IR} = \frac{N}{2}\sum_{n=0}^{\infty} D_n^p \log\left(\frac{\Gamma \left(\frac{d}{2}+n-1\right)}{\Gamma
   \left(-\frac{d}{2}+n+p+1\right)}\right)\,.
\end{equation} 
This sum has exactly the same form as the one corresponding to the change of free energy in a CFT$_d$ perturbed by a double trace operator of dimension $2\Delta$, which was computed in \cite{Diaz:2007an}
\begin{equation}
\begin{aligned}
\delta F &= \frac{1}{2}\sum_{n=0}^{\infty}D_n^d \log\left(\frac{\Gamma \left(n+\Delta\right)}{\Gamma
   \left(n+d-\Delta\right)}\right) \\
&= -\frac{1}{\sin(\frac{\pi d}{2})\Gamma(d+1)}\int_0^{\Delta-\frac{d}{2}}
du\, u \sin(\pi u)\Gamma\left(\frac{d}{2}+u\right)\Gamma\left(\frac{d}{2}-u\right)\,.
\end{aligned}
\end{equation}
Identifying $\Delta = \frac{d}{2}-1$ and $p=d$, we then have for our defect free energy 
\begin{equation}
\cF_{\rm IR} = -\frac{N}{\sin(\frac{\pi p}{2})\Gamma(p+1)}\int_0^{\frac{d}{2}-1-\frac{p}{2}}
du\, u \sin(\pi u)\Gamma\left(\frac{p}{2}+u\right)\Gamma\left(\frac{p}{2}-u\right)\,.
\end{equation}
In the limit $p\rightarrow 2$, this has a pole as expected, reflecting the Weyl anomaly. A convenient way to extract the coefficient of the pole is to follow \cite{Kobayashi:2018lil} and define the quantity $\tilde D_p = -\sin(\frac{\pi p}{2})\cF$. In the $p\rightarrow 2$ limit, this is proportional to the $b$-anomaly coefficient, $\tilde D_{p=2} = \frac{\pi}{6} b$. Then we finally find
\begin{equation}
b_{\rm IR}= 3N\int_0^{\frac{d}{2}-2} du\, u^2 = -N\left(2-\frac{d}{2}\right)^3\,.
\label{b-exact}
\end{equation} 
This is the exact result for the $b$ coefficient at the IR fixed point of the defect RG flow in the free theory. Note that it agrees with (\ref{bpert}) if we let $d=4-\epsilon$. We see that $b_{\rm IR}<0$ in $d<4$, consistent with the $b$-theorem and the fact that the RG flow connects the trivial defect ($b_{\rm UV}=0$) to the fixed point $h=h_*$ in the IR. In $d=3$, (\ref{b-exact}) gives $b_{\rm IR} = -\frac{N}{8}$, which is twice the $b$-anomaly coefficient of $N$ scalars with Dirichlet boundary conditions \cite{Jensen:2015swa}, as expected. For $d=2$, the defect action becomes simply a mass term over the whole space, and so the RG flow should connect $N$ free scalars in the UV to the empty theory in the IR. This is consistent with (\ref{b-exact}), which gives $b_{\rm IR}=-N$ for $d=2$. 

\subsection{Interacting theory in $d=4-\epsilon$}
We now compute the defect free energy in the case of the interacting $O(N)$ model, eq.~(\ref{phi4}). Working up to order $\epsilon^3$, in addition to the diagrams $F_0$, $F_1$ computed above in the free theory case, there is now an additional diagram of order $h^2\lambda$
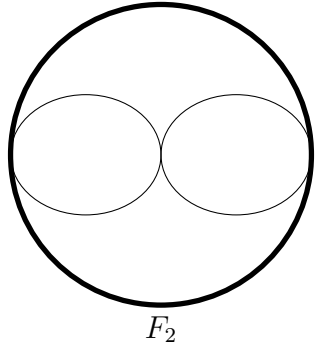
\begin{center}
\begin{tikzpicture}
\coordinate[label=below:{$F_2$}] (A) at (0, -2);
\draw[line width=2] (0, 0) circle (2);
\draw (-1, 0) ellipse (1 and 0.8);
\draw (1, 0) ellipse (1 and 0.8);
\end{tikzpicture}
\end{center}
This can be evaluated as
\begin{equation}
\begin{aligned}
F_2 =& \int_D d^2 x \int_D d^2 y \int d^d z \ (-\lambda_0 h_0^2) \frac{N^2 + 2N}{6} \frac{C_\phi^4}{((z_\parallel-x)^2 + z_\perp^2)^{d-2} ((z_\parallel-y)^2 + z_\perp^2)^{d-2}} \quad\text{eq.~\ref{I1}}\\
=& -\lambda_0 h_0^2 \frac{N^2 + 2N}{6} \int_D d^2 x \int_D d^2 y\ C_\phi^4
\frac{\pi ^{d/2} \Gamma \left(2-\frac{d}{2}\right)^2 \Gamma \left(\frac{3 d}{2}-4\right) |x-y|^{8-3 d}}{\Gamma (4-d) \Gamma (d-2)^2} \\
=& -\lambda_0 h_0^2 \frac{N^2 + 2N}{6} 4\pi R^4 C_\phi^4 \int_0^\pi d\theta \sin(\theta) \int_0^{2\pi} d\phi
\frac{\pi ^{d/2} \Gamma \left(2-\frac{d}{2}\right)^2 \Gamma \left(\frac{3 d}{2}-4\right) |2R\sin(\frac{\theta}{2})|^{8-3 d}}{\Gamma (4-d) \Gamma (d-2)^2} \\
=& \frac{2^{8-5 d} (d-4) \pi ^{5-\frac{3 d}{2}} h_0^2 \lambda _0 N (N+2) R^{-3 (d-4)} \csc ^2\left(\frac{\pi  d}{2}\right) \Gamma \left(\frac{3 (d-4)}{2}\right)}{\Gamma (4-d) \Gamma
   \left(\frac{d-1}{2}\right)^2} 
\end{aligned}
\end{equation}
The defect free energy to this order is then 
\begin{equation}
\cF = - F_0 - F_1 - F_2 + \ldots 
\end{equation}
with $F_0$ and $F_1$ given in (\ref{F0}), (\ref{F1}). Substituting the bare couplings with the renormalized ones as in (\ref{lambda0}) and (\ref{h0}),  expanding to the relevant order in $\epsilon$, and focusing on the $\log(MR)$ dependence, we now find
\begin{equation}
\cF =\left(\frac{N \epsilon h^2}{8\pi^2}-\frac{N h^3}{12\pi^3} -\frac{N(N+2)h^2\lambda}{384\pi^4}\right)\log(MR)+\ldots 
\label{cFlogR-int}
\end{equation}   
At the IR fixed point, we then have
\begin{equation}
b_{\rm IR} = \frac{-3N \epsilon h_*^2}{8\pi^2}+\frac{N h_*^3}{4\pi^3} +\frac{N(N+2)h_*^2\lambda_*}{128\pi^4} = -\frac{27N\epsilon^3}{(N+8)^3}+O(\epsilon^4)\,.
\label{b-interacting}
\end{equation}
This is negative, again consistently with the $b$-theorem (\ref{b-theorem}) for the defect RG flow. 

The result (\ref{b-interacting}) can also be checked by comparing with a general formula derived in \cite{Kobayashi:2018lil} using the conformal perturbation theory approach (this is essentially the same as the standard conformal perturbation theory in a CFT perturbed by a weakly relevant operator, see e.g. \cite{Fei:2015oha}). For a 2-dimensional defect with a perturbing operator of dimension $\Delta = 2-\varepsilon$, the beta function in conformal perturbation theory is $\beta = -\varepsilon h + \pi \frac{C_3}{C_2} h^2+O(h^3)$, with $C_3$ and $C_2$ being the 3-point and 2-point function normalizations (see the discussion in the preceding sections). The change in $\tilde D$ is found to be \cite{Fei:2015oha,Kobayashi:2018lil}
\begin{equation}
\tilde D_{\rm IR}-\tilde D_{\rm UV} = -\frac{\pi}{6} \frac{C_2^3}{C_3^2}\varepsilon^3\,,
\label{tildeD-pert}
\end{equation}
where $\tilde D_{\rm IR} =\tilde D_{h=h^*}$ and $\tilde D_{\rm UV} = \tilde D_{h=0}$ (note that the labels `IR' and `UV' here assume that $h$ is a relevant coupling, so that the non-trivial fixed point is in the IR. This is appropriate in $d<4$). 
To apply this formula to the quartic theory in $d=4-\epsilon$, we note that the perturbing operator $\phi_a^2$ has $\Delta = 2 - \frac{6 \epsilon}{N+8}+O(\epsilon^2)$ at the interacting bulk fixed point $\lambda_*$ and at $h=0$ (the UV fixed point of the defect flow). Hence, we identify $\varepsilon =\frac{6 \epsilon}{N+8}$. The normalization constants are $C_2 = 2NC_{\phi}^2$, $C_3 = 8NC_{\phi}^3$. Then, accounting for the fact that $\tilde D_{\rm UV}=0$ and $\tilde D_{\rm IR} = \frac{\pi}{6}b_{\rm IR}$, we see that (\ref{tildeD-pert}) indeed reproduces (\ref{b-interacting}). 

Note that the conformal perturbation theory formula (\ref{tildeD-pert}) can be also applied directly to the case of the large $N$ expansion in general $d$ (for the case of perturbation theory around the trivial saddle point, as discussed in Section \ref{largeN}). In this case, identifying $\varepsilon=2-\Delta_{\sigma} = -t/N$, one finds
\begin{equation}
b_{\rm IR}^{\rm Large~N}=\frac{6}{\pi}\tilde D_{\rm IR}^{\rm Large~N} = \frac{C_{\sigma}^3}{\left(g_{\sigma^3}\right)^2}\frac{t^3}{N^3}\,.
\end{equation}
Using eqs. (\ref{C_sigma}), (\ref{t}) and (\ref{g_sigma^3}), this gives
\begin{equation}
b_{\rm IR}^{\rm Large~N} = -\frac{(4-d)(d-1)}{d(d-3)^2 N^2}
\left(\frac{2^d (d-2) \sin \left(\frac{\pi  d}{2}\right) \Gamma \left(\frac{d+1}{2}\right)}{\pi ^{3/2} \Gamma \left(\frac{d}{2}+1\right)}\right)^2\,.
\label{largeN-b}
\end{equation}
In $d=4-\epsilon$ and to leading order in $\epsilon$, this agrees with (\ref{b-interacting}). We can also apply (\ref{tildeD-pert}) to the defect RG flow in the cubic theory in $d=6-\epsilon$ (but note that in this case the non-trivial fixed point $h=h^*$ sits in the UV because $h$ is a relevant coupling). Using $\varepsilon = 2 - \Delta_\sigma = \frac{\epsilon}{2} -\frac{1}{(4 \pi)^3} \frac{N\left(g_1^*\right)^2+\left(g_2^*\right)^2}{12}$, as well as $C_2=C_{\phi}$, $C_3 = -g_2^* \pi^3 C_{\phi}^3$, one finds to leading order in small $\epsilon$ and large $N$
\begin{equation}
b_{h=h^*}^{d=6-\epsilon} = \frac{\left(-\frac{\epsilon }{2}+\frac{N(g_1^*)^2 +(g_2^*)^2}{768 \pi ^3}\right)^3}{\pi ^6 (g_2^*)^2 C_{\phi }^3} = 
\frac{8000\epsilon^2}{27N^2}+\ldots 
\end{equation}
This can be see to agree with the large $N$ result (\ref{largeN-b}) expanded in $d=6-\epsilon$ (keep in mind that the label `IR' should be switched to `UV' in this case since it is the trivial defect that sits in the IR). 

Let us also briefly comment on the renormalized defect entropy function recently proposed in \cite{Shachar:2022fqk} as a monotonically decreasing function along the flow. This is defined as \cite{Shachar:2022fqk}
\begin{equation}
{\cal S} =\frac{1}{2}\left(R^2 \partial_R^2-R \partial_R\right) \mathcal{F}\,.
\end{equation}
The differential operator acting on $\cF$ is such that at the fixed points, the entropy function ${\cal S}$ is proportional to the $b$ anomaly coefficient, namely ${\cal S}=b/3$, see eq.~(\ref{Fvsb}). From the above perturbative evaluation of $\cF$ in the interacting theory in $d=4-\epsilon$, we find
\begin{equation}
{\cal S}=-\frac{N \epsilon h^2}{8\pi^2}+\frac{N h^3}{12\pi^3}+\frac{N(N+2)h^2\lambda_*}{384\pi^4} = 
-\frac{3 N \epsilon h^2}{4 (N+8)\pi^2}+\frac{h^3 N}{12 \pi ^3}
\label{entropy}
\end{equation}
where we have plugged in the fixed point value of the bulk coupling, but kept $h$ general as the entropy function can be defined along the defect RG flow. It is straightforward to check that $\partial {\cal S}/\partial h <0$ along the flow from $h=0$ to $h=h_* = 6\pi \epsilon/(N+8)$. Note also that $\partial {\cal S}/\partial h$ is simply proportional to the beta function (\ref{betah:lambdah}). The entropy function can be similarly computed for the defect RG flow in the free theory, which gives ${\cal S}=-\frac{N \epsilon h^2}{8\pi^2}+\frac{N h^3}{12\pi^3}$, and $\partial {\cal S}/\partial h = N\beta_h/(4\pi^2)$ is negative along the flow. 

\section{Conclusions} 
\label{Sec:Conclusions}
We studied the critical behavior of a surface defect in the $O(N)$ model, using both epsilon and large $N$ expansions, finding evidence that the system flows to a non-trivial DCFT at low energies. We also computed the spherical defect free energy and related  Weyl anomaly $b$-coefficient, and checked consistency with the $b$-theorem. 

For future work, one obvious direction is to extend our calculations to higher orders in $\epsilon$ and $1/N$, and perhaps make contact with the numerical conformal bootstrap or Monte Carlo simulations. In this note we focused on the scaling dimensions of the simplest defect operators, but it would be interesting to study the DCFT data in more detail. For example, it would be useful to analyze in depth the 2-point function of bulk scalars in the presence of the defect (for instance using the equations of motion method as in \cite{Giombi:2020rmc, Giombi:2022vnz}), and extract the relevant OPE data.  

In our analysis of the large $N$ expansion in Section \ref{sec:largeN}, we focused on perturbation theory around the trivial saddle point with $\sigma=0$. It would be useful to study the $\sigma$ effective action in detail and see if there are non-trivial ``classical" saddle points with $\sigma \sim \sqrt{N}$. As explained in Section \ref{largeN} above, we expect at least one such saddle point corresponding to the $O(N)$ breaking phase, but there may be other $O(N)$ invariant saddle points (similarly to the boundary case analyzed in \cite{Giombi:2020rmc}). This may be relevant to elucidate the transition between the large $N$ behavior in $3<d<4$ and that in $d=3$.  

Another direction is to study similar surface defects in other theories, such as the Gross-Neveu CFT. In this case, the fermion bilinear $\bar\psi \psi \sim \sigma $ has dimension slightly below 1 at large $N$, so it is natural to consider a surface defect with action $S_D \sim \int d^2 x \sigma^2(\vec{x},0)$. A similar defect can be considered in the Gross-Neveu-Yukawa description near $d=4$, where $\sigma$ becomes a propagating scalar field, so the surface operator would be analogous to the one studied in this paper in the interacting scalar theory in $d=4-\epsilon$. 

It would be also interesting to see if the surface defect has a natural holographic description. The free and critical $O(N)$ models in $d$-dimensions, restricted to the $O(N)$ singlet sector, are dual to the Vasiliev higher spin theory in AdS$_{d+1}$ (the free and critical theory correspond to alternate choice of boundary conditions for the bulk scalar field dual to the $\phi_a^2$ operator, see \cite{Giombi:2016ejx} for a review). Since the surface defect (\ref{phi4-defect}) is defined in terms of the $O(N)$ invariant ``single trace" operator $\phi_a^2$ dual to the bulk scalar field, it should have a realization in the higher spin theory. A natural starting point would be to consider a $H^3\times S^{d-3}$ slicing of AdS$_{d+1}$, and look for configurations where the metric and bulk scalar (and possibly the higher spin fields) have a non-trivial profile preserving the isometries of $H^3\times S^{d-3}$ (since these are the symmetries of the DCFT). It would be interesting to explore this and make contact with the predictions of the large $N$ expansion. 

\section*{Acknowledgments}
We thank Gabriel Cuomo and Max Metlitski for useful discussions, and Avia Raviv-Moshe and Siwei Zhong for sharing a draft of their work. The research of SG and BL is supported in part by the US NSF under Grant No.~PHY-2209997.

\appendix
\section{Formulas}
In this appendix we collect some useful formulas that were used for the calculations in the main text of the paper. 

{Fourier transform:}
\begin{equation}\label{fourier}
\int d^{d} x \frac{e^{-i k x}}{x^{a}}
= \frac{2^{d-a} \pi^{d / 2} \Gamma((d-a) / 2)}{\Gamma(a / 2)} \frac{1}{k^{d-a}} 
\end{equation}
Eq.~\ref{I1}, \ref{I3}, and \ref{I38} can be found in~\cite{Smirnov:2006ry}; eq.~\ref{2prop} in~\cite{kleinert2001critical}:
\begin{equation}
\begin{aligned}
\int \frac{d^{d} q}{(2 \pi)^{d}} \frac{1}{q^{2 \alpha}(p+q)^{2 \beta}} 
=\frac{1}{(4 \pi)^{\frac{d}{2}}} \frac{\Gamma\left(\frac{d}{2}-\alpha\right) \Gamma\left(\frac{d}{2}-\beta\right) \Gamma\left(\alpha+\beta-\frac{d}{2}\right)}{\Gamma(\alpha) \Gamma(\beta) \Gamma(d-\alpha-\beta)}\left(\frac{1}{p^{2}}\right)^{\alpha+\beta-\frac{d}{2}}
\end{aligned}
\label{I1}
\end{equation}
\begin{equation}
\label{I3}\begin{aligned}
\int \frac{\mathrm{d}^{d} k}{(2\pi)^d} \frac{1}{\left(k^{2}+m^{2}\right)^{\lambda_{1}}\left(k^{2}\right)^{\lambda_{2}}} 
= 
\frac{\Gamma\left(\lambda_{1}+\lambda_{2}-d/2\right) \Gamma\left(-\lambda_{2}+d/2\right)}
{(4\pi)^{d/2} \Gamma\left(\lambda_{1}\right) \Gamma(d/2)} \frac{1}{\left(m^{2}\right)^{\lambda_{1}+\lambda_{2}-d/2}}
\end{aligned}
\end{equation}
\begin{equation}\label{I38}
\begin{aligned}
&\iint \frac{\mathrm{d}^{d} k \mathrm{~d}^{d} l}{\left(k^{2}+m^{2}\right)^{\lambda_{1}}\left[(k+l)^{2}\right]^{\lambda_{2}}\left(l^{2}+m^{2}\right)^{\lambda_{3}}} \\
=& \frac{\pi^{d} \Gamma\left(\lambda_{1}+\lambda_{2}-d/2\right) \Gamma\left(\lambda_{2}+\lambda_{3}-d/2\right) \Gamma\left(d/2-\lambda_{2}\right) \Gamma\left(\lambda_{1}+\lambda_{2}+\lambda_{3}-d\right)}
{\Gamma\left(\lambda_{1}\right) \Gamma\left(\lambda_{3}\right) \Gamma\left(\lambda_{1}+2 \lambda_{2}+\lambda_{3}-d\right) \Gamma(d/2)\left(m^{2}\right)^{\lambda_{1}+\lambda_{2}+\lambda_{3}-d}} 
\end{aligned}
\end{equation}
\begin{equation}
\begin{gathered}
\int \frac{d^{D} p}{(2 \pi)^{D}} \frac{1}{\left(\mathbf{p}^{2}+m^{2}\right)^{a}\left[(\mathbf{p}-\mathbf{k})^{2}+m^{2}\right]^{b}} 
= \frac{\Gamma(a+b{-D/2})}{(4 \pi)^{D / 2} \Gamma(a) \Gamma(b)} \int_{0}^{1} d x \frac{(1-x)^{a-1} x^{b-1}}{\left[m^{2}+\mathbf{k}^{2} x(1-x)\right]^{a+b-D / 2}} 
\end{gathered}\label{2prop}
\end{equation}

For {$a_{1}+a_{2}+a_{3}=d$}, the uniqueness relation is \cite{Chai:2021uhv}
\begin{equation}
\int d^{d} x \frac{1}{\left|x_{1}-x\right|^{2 a_{1}}\left|x_{2}-x\right|^{2 a_{2}}\left|x_{3}-x\right|^{2 a_{3}}}=\frac{U\left(a_{1}, a_{2}, a_{3}\right)}{\left|x_{12}\right|^{d-2 a_{3}}\left|x_{13}\right|^{d-2 a_{2}}\left|x_{23}\right|^{d-2 a_{1}}}
\label{triangle}\end{equation}
where
\begin{equation}
U(a, b, c) 
=\frac{\pi^{\frac{d}{2}} \Gamma\left(\frac{d}{2}-a\right) \Gamma\left(\frac{d}{2}-b\right) \Gamma\left(\frac{d}{2}-c\right)}{\Gamma(a) \Gamma(b) \Gamma(c)}
\end{equation}

\bibliographystyle{ssg}
\bibliography{2dDefectDraft-bib}

\end{document}